\documentclass[aps,prb,twocolumn,showpacs,superscriptaddress,10pt]{revtex4-1}
\usepackage{amsmath}
\usepackage{amsfonts}
\usepackage{amssymb}
\usepackage{graphicx}
\usepackage{color}
\usepackage[normalem]{ulem}

\newcommand{\rr}{\mathbf{r}}

\newcommand{\ee}{\mathrm{e}}

\newcommand{\qq}{\mathbf{q}}
\newcommand{\pp}{\mathbf{p}}

\newcommand{\SP}{\mathbf{S}}
\newcommand{\B}{\mathcal{B}}
\newcommand{\A}{\mathcal{A}}

\newcommand{\newtext}[1]{#1}

\begin{document}

\title{Cavity-induced quantum spin liquids}

\author{Alessio Chiocchetta}
\email{Corresponding author. Email: achiocch@uni-koeln.de}
\affiliation{Institute for Theoretical Physics, University of Cologne, D-50937 Cologne, Germany}
\author{Dominik Kiese}
\affiliation{Institute for Theoretical Physics, University of Cologne, D-50937 Cologne, Germany}
\author{Carl Philipp Zelle}
\affiliation{Institute for Theoretical Physics, University of Cologne, D-50937 Cologne, Germany}
\author{Francesco Piazza}
\affiliation{Max Planck Institute for the Physics of Complex Systems, 01187, Dresden, Germany.}
\author{Sebastian Diehl}
\affiliation{Institute for Theoretical Physics, University of Cologne, D-50937 Cologne, Germany}

\date{\today}

\maketitle

\textbf{Quantum spin liquids provide paradigmatic examples of highly entangled quantum states of matter. Frustration is the key mechanism to favor spin liquids over more conventional magnetically ordered states. Here we propose to engineer frustration by exploiting the coupling of quantum magnets to the quantized light of an optical cavity. The interplay between the quantum fluctuations of the electro-magnetic field and the strongly correlated electrons results in a tunable long-range interaction between localized spins. This cavity-induced frustration robustly stabilizes spin liquid states, which occupy an extensive region in the phase diagram spanned by the range and strength of the tailored interaction. Remarkably, this occurs even in originally unfrustrated systems, as we showcase for the Heisenberg model on the square lattice.}

Quantum spin liquids (QSLs) represent strongly correlated phases of matter, which are characterized by quantum fluctuations so dominant as to suppress magnetic ordering down to the lowest temperatures. Yet, the spins may be quantum mechanically entangled over long distances~\cite{Balents_review,Ng_review,Moessner_review}. In Nature, QSLs are expected to occur in proximity to magnetic phases, but their existence often remains elusive. The key ingredient behind quantum spin liquid formation is, however, clearly identified: it is the presence of strong frustration, which disallows magnetic symmetry breaking, but need not be averse to, e.g. quantum mechanical singlet ordering. The routes towards frustration are manifold: one promising avenue is the focus on materials where magnetic ordering is penalized by the geometry of the lattice, such as for triangular, Kagom\'{e} or pyrochlore lattices~\cite{Balents_frustration,Castelnovo2012,Gingras2014,Norman2016}. Another one proceeds via the energetic competition of couplings of different range, like in the antiferromagnetic $J_1-J_2$ Heisenberg model or dipolar-interacting systems~\cite{Yao2018,Zou2017,Keles2018},
where the simultaneous appearance of nearest- and beyond-nearest-neighbour couplings counteracts global antiferromagnetism.

The challenge is then out to engineer robust QSL states of quantum condensed matter. Here, we will achieve this task by coupling an ordinary Heisenberg antiferromagnet on a square lattice to the electro-magnetic field of an optical cavity. 

\begin{figure*}[t!]
\includegraphics[width=11.1cm]{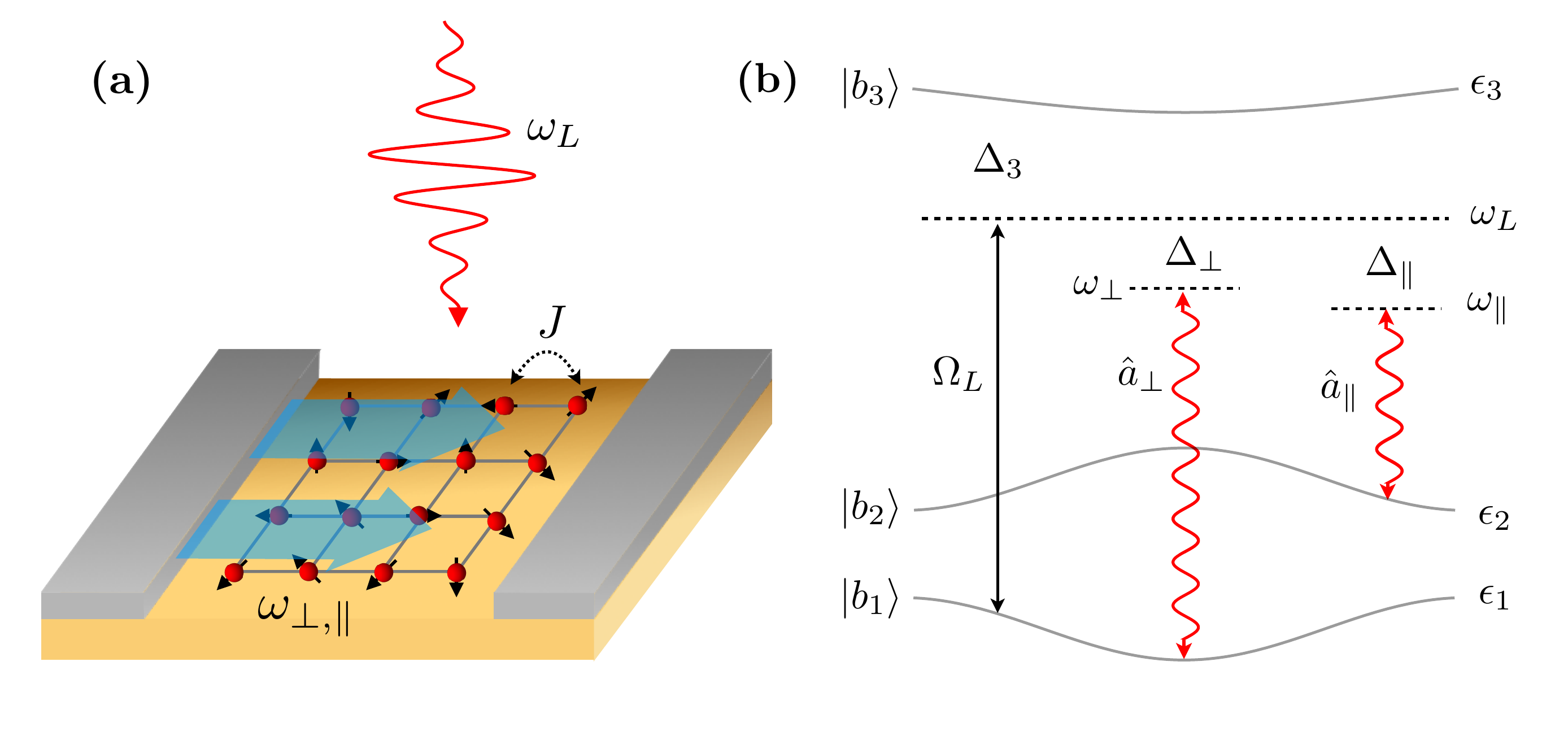}
\includegraphics[width=6.72cm]{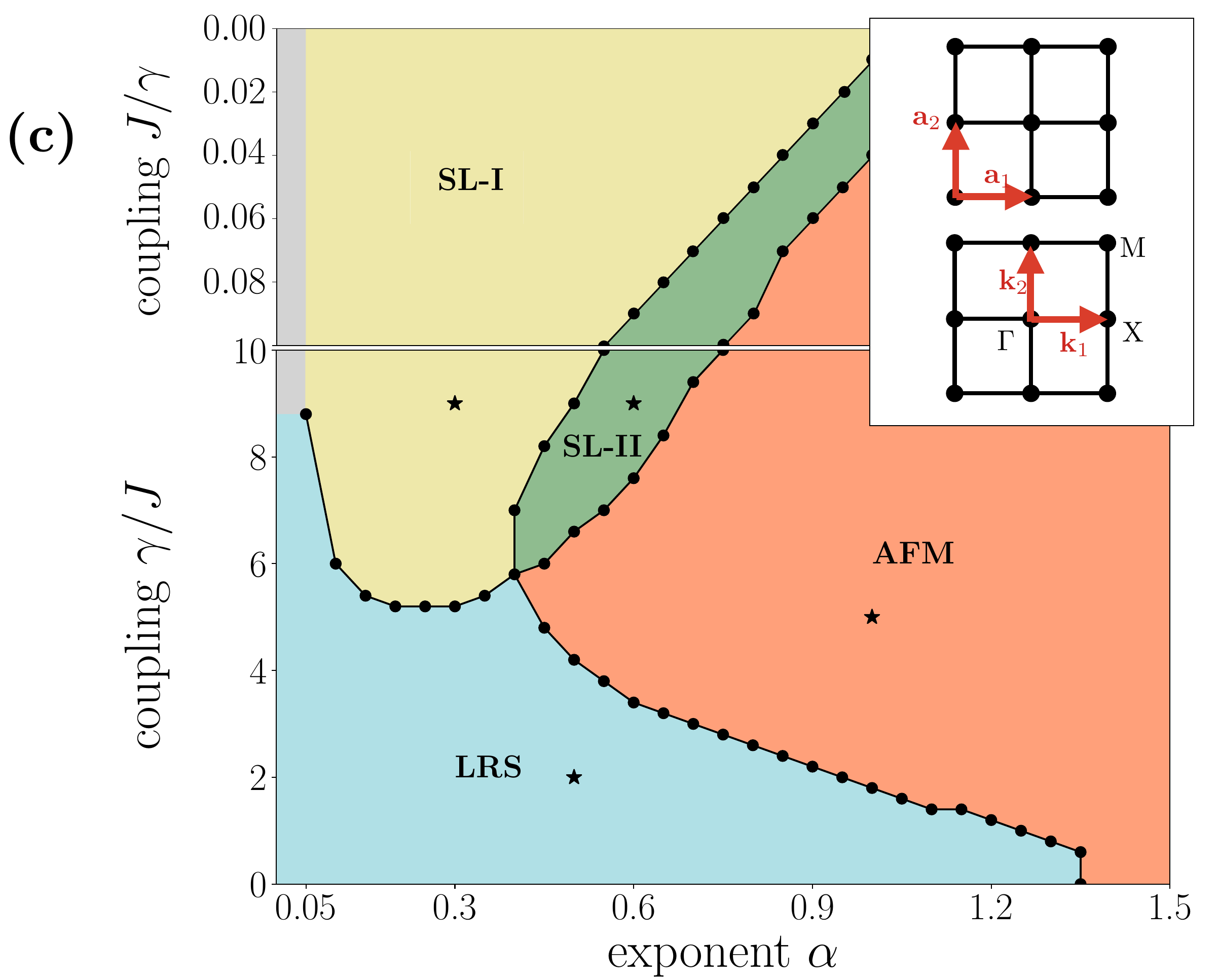}
\caption{\textbf{Implementation of a cavity-induced quantum spin liquid and phase diagram.} 
\textbf{(a)} Setup: a two-dimensional material, with nearest-neighbour exchange interaction $J$, is coupled to a cavity with fundamental frequencies $\omega_{\perp,\parallel}$, whose field is represented by the light-blue arrows. The system is driven by an external laser with frequency $\omega_L$.
\textbf{(b)} Level scheme: the electronic orbitals $|b_{1,2}\rangle$, with energies $\sim\epsilon_{1,2}$ are coupled to the auxiliary band $|b_3\rangle$, with energy $\sim \epsilon_3$ via the laser with Rabi frequency $\Omega_L$ and the cavity modes $ {a}_{\perp, \parallel}$. The third band is detuned from the laser by $\Delta_3$, and from the cavity modes by $\Delta_{\perp,\parallel}$.
\textbf{(c)} Phase diagram for the ground state of Hamiltonian~\eqref{eq:Hamiltonian}, obtained from the bosonic spinon decomposition, as a function of the exponent $\alpha$ and of the coupling ratio $\gamma/J$. \newtext{Error bars on the phase boundaries are within the symbols' size.} The inset shows the square lattice and the reciprocal one, with the respective primitive vectors. Discussion and abbreviations, see text.
}
\label{fig:fig1}
\end{figure*}

The physical mechanism stabilizing the QSL takes the second route towards strong frustration to the extreme, by considering long-range antiferromagnetic interactions described by an algebraically decaying spin-spin interaction $\sim r^{-\alpha}$ including the case of all-to-all couplings $\alpha =0$, mediated by the cavity, cf.~Fig.~\ref{fig:fig1}  {(a)}. For the limiting case $\alpha =0$ and a cavity induced interaction $\gamma$ dominating over the nearest-neighbour Heisenberg coupling $J$, $J/\gamma=0$, this realizes a \newtext{ state with long-range correlations mediated by  singlets of arbitrarily large size (LRS) }. Away from this limit, and for decay exponents $\alpha \lesssim 1$, within a Schwinger-boson approach  we find that the frustration imprinted by the cavity creates an extensive regime of QSL states. It is characterized by the absence of spontaneous symmetry breaking, and fractional excitations of both of a gapped (SL-I) and of gapless (SL-II) nature, cf. Fig.~\ref{fig:fig1}  {(b)}. As a consequence of the underlying long-ranged interactions, correlations decay algebraically in both these phases. 

In terms of a physical implementation, we draw motivation from recent developments exploring the interplay of quantum materials with quantized light. This idea has been researched in the context of weakly correlated systems, mainly as a tool to reinforce superconductivity and other coherent many-body phases~\cite{Sentef2018,Mazza2019,Andolina2019,Schlawin2019,Curtis2019,Thomas2019,Gao2020,ashida2020,schuler2020,chakraborty2020}. First works have also addressed the strong coupling regime, showing how existing phases can be manipulated in this way~\cite{ Sentef2020,Kiffner2019,Mazza2019}. Here we demonstrate that the coupling to a cavity can even induce phases that are not present in its absence: an unfrustrated antiferromagnetic system is turned into a quantum spin liquid, provided the antiferromagnetic interaction mediated by the cavity is sufficiently long-ranged and strong. To achieve these requirements, we develop a solid-state implementation harnessing localized electronic orbitals as effective spin degrees of freedom, coupled to the cavity modes via additional coherent laser drive, cf. Fig.~\ref{fig:fig1}~{(a)}. This gives rise to quantum mechanically fluctuating, effective magnetic fields in all linearly independent spatial directions, which vanish on average. They thus counteract dynamically magnetization in any direction, but do not suppress the spin-singlet ordering, crucial for QSL states.

%
We consider a long-range SU(2)-symmetric Heisenberg model on a square lattice
\begin{equation}
\label{eq:Hamiltonian}
H = J\sum_{\langle i,j \rangle}\,  \SP_i \cdot \SP_j + \gamma \sum_{ i \neq j }\, \frac{\SP_i \cdot \SP_j}{|\rr_{ij}|^\alpha} ,
\end{equation}
with $\SP_i = (S^x_i, S^y_i,S^z_i)$ spin-$1/2$ operators on the lattice site $i$, $J>0$ the nearest-neighbour antiferromagnetic exchange, $\gamma > 0$ the strength of the long-range interaction modulated by the exponent $\alpha$, and $\rr_{ij} \equiv \rr_i-\rr_j$. \newtext{Periodic boundary conditions are assumed.}
Before analysing the ground-state phase diagram of the Hamiltonian~\eqref{eq:Hamiltonian}, let us qualitatively discuss the expected phases, starting with some known limiting cases.
For $\gamma = 0$, the ground state of the Hamiltonian~\eqref{eq:Hamiltonian} displays N\'{e}el-like order~\cite{Auerbach_book}. 
For $\alpha = 0$ and $\gamma\gg J$,  the long-range Hamiltonian is proportional to the total spin  $(\sum_i \SP_i)^2$: this imposes a constraint on this singlet manifold, energetically penalizing states with a finite value of the total spin $S$, including states with finite magnetization. As a result, the ground state of the total Hamiltonian is given by the ground state of the short-range Hamiltonian projected on the singlet manifold. 
\newtext{This is similar to the analysis in Ref.~\onlinecite{Liang1988}, where RVB states with singlets of arbitrarily large size were used as variational wavefunctions. We will denote this state as a long-range singlet state (LRS). }
Finally, for $J=0$, different scenarios are possible: for $\alpha$ large enough, only nearest-neighbouring sites experience an appreciable interaction, and therefore  N\'{e}el-like order is expected. For smaller values of $\alpha$, the frustrating nature of the interaction is expected to penalize AFM order, thus favoring disordered phases. This was shown to be the case for $\alpha =3$ on the triangular lattice~\cite{Yao2018}, and on the square lattice~\cite{Zou2017} (although only \newtext{for spatially anisotropic interactions} in the latter case), where a QSL phase was found. 

Summarizing, by varying $\gamma/J$ and $\alpha$, we expect three kinds of phases: (i) N\'{e}el-like AFM, (ii) a disordered QSL phase, and (iii) a \newtext{LRS} phase. This is substantiated below using a Schwinger-boson approach, which is capable of capturing all the phases mentioned above. In particular, it provides a natural interpolation scheme between the well-understood RVB and N\'{e}el physics discussed above. 

\begin{table}[t]
\begin{ruledtabular}
\begin{tabular}{lllll}
  		& {SL-I}	& {SL-II}	& {\newtext{LRS}}		& {AFM}\\  \hline
Gap  &  Yes		 		& No					& Yes 			&	No			 \\  
LRO 	&  No   				& No					& Yes			& Yes			   \\    

\end{tabular}
 \caption{\textbf{Ground-state phases}. Summary of the four phases identified in this work, according to the criteria discussed in the main text.}
\label{tab:tab1}
\end{ruledtabular}
\end{table}

\begin{figure*}[t!]
\includegraphics[width=\textwidth]{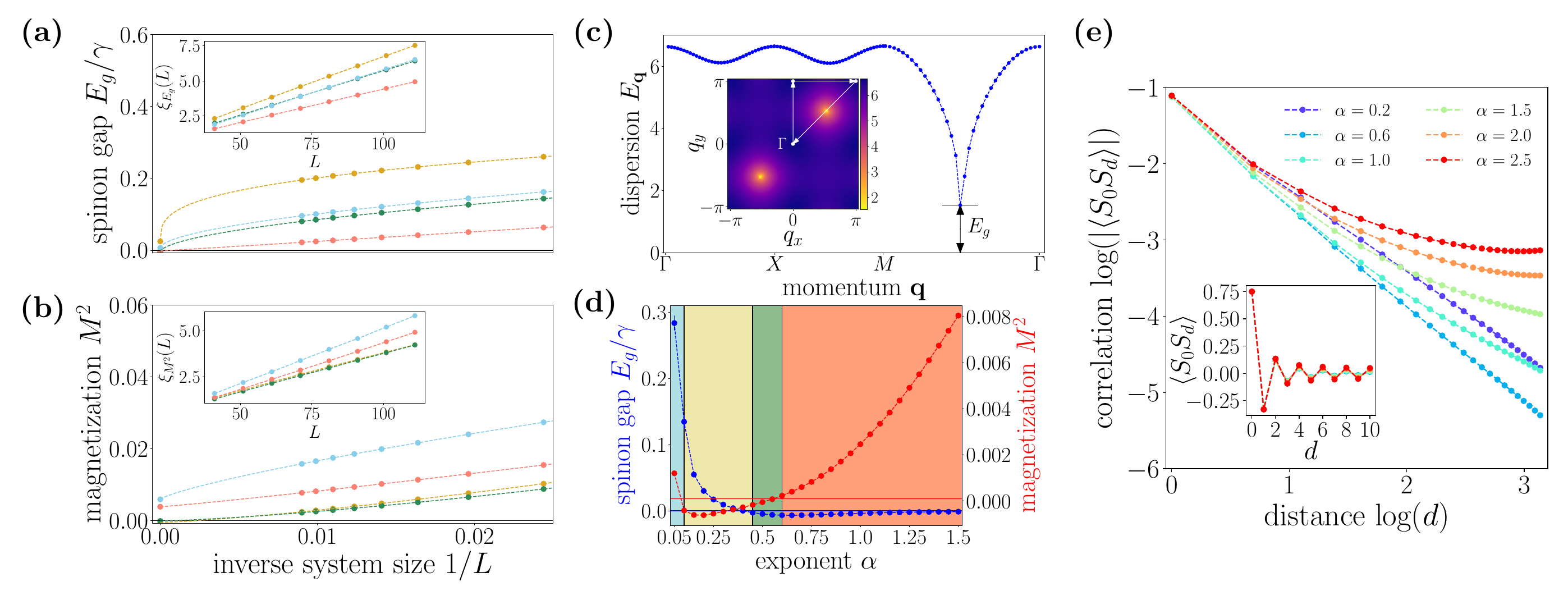}
\caption{\textbf{Numerical characterization of the ground state.} 
\textbf{(a)} and~\textbf{(b)}: dependence of the spinon gap (upper panel) and  square magnetization (lower panel) on the inverse linear system size. The curves refer to values of $\alpha$ and $\gamma/J$ denoted in Fig.~\ref{fig:fig1}~\textbf{(c)} by star symbols, according to the corresponding background colors. The maximum linear size considered is $L=110$. Insets: values of the functions $\xi_{E_g, M^2}(L)$ as functions of the linear system size $L$. 
\textbf{(c)}: spinon dispersion for $\gamma = 7J$ and $\alpha = 0.3$ (SL-I phase), for given cuts in the first Brillouin zone. Inset: spinon dispersion in the first Brillouin zone. The white lines denote the cuts of the main plot. 
\text{(d)}: extrapolated gap (blue curve) and square magnetization (red curve) as functions of the exponent $\alpha$. The background colors reflect the phases illustrated in Fig.~\ref{fig:fig1}~\textbf{(c)}. 
\textbf{(e)}: spin-spin correlation functions along the lattice axis for different values of the exponent $\alpha$ and for $\gamma = 7J$ Inset: spin-spin correlations at short distances. 
}
\label{fig:fig2}
\end{figure*}

%

In order to unveil the nature of the ground state of the Hamiltonian~\eqref{eq:Hamiltonian}, we apply the bosonic spinon decomposition pioneered in Refs.~\onlinecite{Arovas1988,Read1991,Ceccatto1993}, where the spin operators are represented in terms of new bosonic degrees of freedom, ultimately interpreted as emergent fractional excitations.
The main advantage of this method is its flexibility to interpolate between the different states previously identified. 
On the one hand, SU(2)-symmetric bosonic ground states are identified with spin liquids. On the other hand, the onset of magnetic order is signalled by the Bose-Einstein condensation of these bosons.  

The spin operators on the lattice site $j$ are decomposed as (using sum convention for the Greek indices)
\begin{equation}
\mathbf{S}_{j}=\frac{1}{2}b_{j ,\mu}^{\dagger} \boldsymbol{\sigma}_{\mu \nu} b_{j, \nu}, 
\end{equation}
where $b_{j, \mu}$ is a boson (spinon) with spin $\mu \in \{ \uparrow,\downarrow\}$, and $\boldsymbol{\sigma}$ the vector of Pauli matrices. The mapping is then completed by the constraint $ b_{j\mu}^\dagger b_{j\mu} = 1$.
Insights on the nature of the state are then obtained from the expectation values of the SU(2)-invariant bilinears $\A_{ij}~=~i \sigma^y_{\mu\nu} \langle b_{i\mu}b_{j\nu}\rangle /2,$ and $\B_{ij} =  \langle b^\dagger_{i\mu}b_{j\mu}\rangle /2$, which indicate the tendency of the spins at the sites $i$ and $j$ of forming a singlet or to align, respectively. For SU(2)-symmetric states, finite values of $\A_{ij}$ and $\B_{ij}$ determine a finite spinon hopping between the lattice sites $i$ and $j$, thus signalling the emergence of propagating fractional excitations. 

After performing a mean-field decoupling of the spinonic Hamiltonian (see Methods for further details), the values of $\A_{ij}$ and $\B_{ij}$ are self-consistently determined by minimizing the ground-state energy. This task is enabled in practice by using an Ansatz for the values of $\A_{ij}$ and $\B_{ij}$. 
The most natural choice is the manifestly translational-invariant Ansatz $\A_{ij} =\A_{i-j}$, $\B_{ij} =\B_{i-j}$, which follows from a projective-symmetry-group analysis~\cite{Wen2002}. 
The resulting saddle-point equations, reported in Eq.~\eqref{eq:HF-equations}, are reduced to a system of $2N + 1$ coupled non-linear equations, for finite-size systems with $N = L \times L$ lattice sites. The numerical complexity of the problem still limits the size $N$ of the systems for which a solution can be found.

For finite-size systems a spontaneous symmetry breaking cannot occur, and therefore the AFM order parameter always vanishes. 
Accordingly, other criteria are needed to assess the onset of an ordered phase. Here we identify the onset of an AF ordered phase when the two following conditions are met: (i) the gap $E_g \equiv \min_\qq E_\qq$ in the spinon dispersion closes upon increasing the system size $N$ and (ii) the squared magnetization $M^2~\equiv~\sum_{j} |\SP_j \cdot \SP_0|/N$ approaches a constant value upon increasing $N$.  Notice that these two indicators also naturally lend themselves to characterize the other phases outlined before: a phase with  $M^2=0$ corresponds to either  a gapped ($E_g\neq 0$) or a gapless $E_g = 0$ QSL, while a phase with $M^2\neq 0$ and $E_g\neq 0$ can be naturally identified with a \newtext{LRS} state. These criteria are summarized in Tab.~\ref{tab:tab1}. 

Let us finally discuss the phase diagram in Fig.~\ref{fig:fig1}  {(c)}. The first, main result, is the emergence of a gapped QSL phase (denoted as SL-I) for $\alpha \lesssim 1.25 $, and $\gamma \gtrsim 5J$, characterized by the presence of a gap and by the absence of long-range correlations. This phase appears for any $\alpha>0.05$, corresponding to the minimum value here considered, suggesting that the \newtext{LRS} phase is unstable in this region and only exists for $\alpha=0$. Additionally, our data also show the existence of a gapless QSL phase (denoted SL-II) for intermediate values of $\alpha$, clearly manifested in the largest available system sizes, as shown in Fig.~\ref{fig:fig2}~{(a)} and~{(b)}.

For $\gamma \lesssim 5J$, the \newtext{LRS} phase is remarkably stable for $\alpha \lesssim 1.25 $. Here, the system is simultaneously gapped and characterized by long-range correlations (cf. Tab.~\ref{tab:tab1}), which, however, do not correspond to a spontaneous symmetry breaking.  Finally, we observe that, as expected, for large values of $\alpha$, as well as for $\gamma=0$, the system is always in the ordinary N\'{e}el-AFM phase.

An example of extrapolated values of $M^2$ and $E_g$ used to build the phase diagram in Fig.~\ref{fig:fig1}  {(c)} is shown in Fig.~\ref{fig:fig2}~{(d)}, as a function of $\alpha$ for $\gamma= 7 J$. 
The fitting function used to extrapolate the $L\to \infty$ limit of these observables has the form $\mathcal{O}_L = \mathcal{O}_\infty + b_\mathcal{O} L^{-\omega_\mathcal{O}}$, with $\mathcal{O}_\infty$, $b_\mathcal{O}$, and $\omega_\mathcal{O}$ fitting parameters.
\newtext{The slightly negative extrapolated values of $M^2$ and $E_g$ are due to the simplified form of the extrapolation function above, which neglects subleading terms in $1/L$ (cf. Ref.~\onlinecite{Sorella}). }
This fitting function was identified by a preliminary evaluation of the quantity $\xi_{\mathcal{O}}(L)=1/\ln \left(\frac{\mathcal{O}_{L-4}-\mathcal{O}_{L-2}}{\mathcal{O}_{L-2}-\mathcal{O}_{L}}\right)$, which displays a linear behaviour in $L$ for algebraic finite-size scaling, while it saturates for an exponential one~\cite{Golinelli1994}. The algebraic finite-size scaling occurring also for gapped phases is imprinted by the algebraic character of the interactions~\cite{Ruffo_book}. 
For the same reason, the spin-spin correlation functions in the QSL phases also display an algebraically decaying behaviour, rather than the usual short-range one, with an exponent depending continuously on the interaction's exponent $\alpha$ (cf. Fig.~\ref{fig:fig2}  {(e)}). Algebraic correlations were similarly found for gapped, disordered phases in spin chains with long-range interactions~\cite{Hauke_2010,Koffel2012,Peter2012}, further substantiating the generality of this mechanism.

\begin{figure}[t!]
\includegraphics[width=0.45\textwidth]{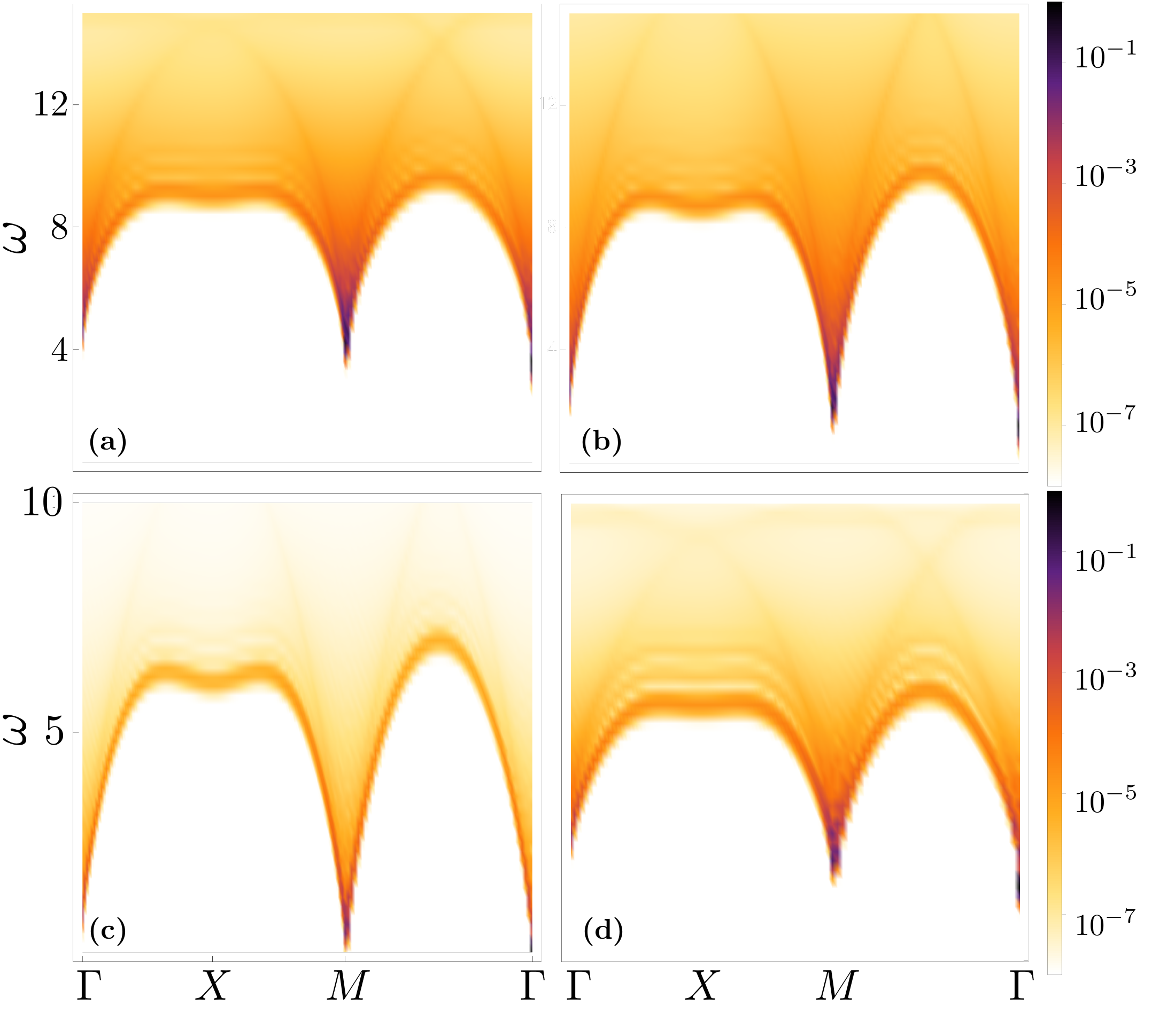}
\newtext{\caption{\textbf{Dynamical structure factor.} $S_\qq(\omega)$  as a function of the frequency $\omega$ and of the momentum $\qq$. Results are shown for: SL-I [panel (a), $\gamma = 9J, \alpha = 0.3$] , SL-II [panel (b), $\gamma=9J, \alpha = 0.6$], AFM [panel (c), $\gamma = 5 J, \alpha = 1.0$], LRS [panel (d), $\gamma = 4.4J, \alpha= 0.35$]. }
}
\label{fig:DSF}
\end{figure}

\newtext{Besides gap and long-range order, we provide a further observable to characterize the phases here identified, i.e., the dynamical structure factor $S_\qq(\omega) = \int_t e^{i \omega t}\left\langle \mathbf{S}_{-\mathbf{q}}(t) \cdot \mathbf{S}_{\mathbf{q}}(0)\right\rangle$, with $\mathbf{S}_{\mathbf{q}}$ the Fourier transform of the spin operators with momentum $\qq$.  $S_\qq(\omega)$, which can be straightforwardly computed from the spinon decomposition~\cite{Messio2010}, leads to markedly different features depending on the phase. 
For the SL-I and SL-II phases (Fig.~\ref{fig:DSF} (a) and (b), repsectively), the DSF features a broadening originated in the continuum of fractional excitations. On the contrary, the AFM phase  (Fig.~\ref{fig:DSF} (c)) shows a sharper signal close to the gapless quasi-particle dispersion, corresponding to the magnonic dispersion expected in the AFM phase. Finally, the LRS phase (Fig.~\ref{fig:DSF} (d))  features a broadening similar to the SL-I phase, suggesting the presence of fractionalized excitations.
}

\newtext{A final word of caution concerns the accuracy of the bosonic spinon decomposition used here. As a mean-field theory, it provides a qualitatively correct topology of the phase diagram, while the phase borders cannot be expected to be quantitatively accurate. }

%
%

\newtext{The Hamiltonian Eq.~\eqref{eq:Hamiltonian} (or variations of it) can be realized in quantum simulators using trapped ions or ultracold atoms~\cite{richerme2014,Yao2018,mivehvar2021cavity}. While these platforms provide unprecedented controllability, the realization of low-temperature strongly correlated phases remains challenging. On the converse, solid-state platforms naturally feature strongly-correlated physics at cryogenically accessible temperatures. Moreover, the controllability in 2D materials is progressing fast, making them, among others, candidates for quantum simulators~\cite{Kennes2021}. In the following, we will focus on a scheme for implementing the Hamiltonian~\eqref{eq:Hamiltonian} in a solid-state system.}

\newtext{Our proposal uses} two
electronic orbital degrees of freedom, constituting a pseudospin of length $S=1/2$. 
In the absence of a cavity, the pseudospins are assumed to be described by a short-range antiferromagnetic Heisenberg model, emerging as a strong Mott limit of a Hubbard model for the electronic degrees of freedom: this is the case, e.g., of iridates and ruthenates materials~\cite{Jackeli2009,Georges2013,Rau2016}. We assume SU(2) symmetry for the sake of simplicity.

As substantiated further below, the coupling of the localized electronic states to the cavity will result in a coupling between the pseudospins and quantized effective magnetic fields. The setup we consider is sketched in Fig.~\ref{fig:fig1}~{(a)}. Two aspects of the long-range Hamiltonian~\eqref{eq:Hamiltonian} are essential to unveil QSL phases: (i) an antiferromagnetic character of the induced interaction and (ii) a high degree of symmetry, ideally SU(2). In order to control the symmetry of the emerging cavity-mediated interaction, we propose to use two cavity modes. While a single mode is sufficient to mediate a U(1)-symmetric interaction, a second mode allows for an enhancement to SU(2) symmetry. The required selectivity in the cavity-spin coupling can be achieved via an auxiliary third band which is driven far off-resonance by a laser [see Fig.~\ref{fig:fig1}~{(b)}]. The resulting two-photon transitions involve virtual excitations to the third band and back to one of the two bands implementing the pseudospin degree of freedom. The sign of the cavity-mediated interaction is then finally determined by the detuning between the laser and each cavity mode.

The paramagnetic and diamagnetic coupling terms between electrons and electromagnetic field are given by:
\begin{equation}
  \label{eq:int_ham}
H_{\rm int}=\frac{1}{2m}\int_\rr
  \psi^\dag(\mathbf{r}) \big [2e \,
  \mathbf{p}\cdot\mathbf{A}(\mathbf{r},t) + e^2 \mathbf{A}^2(\mathbf{r},t)\bigg]\psi(\mathbf{r})\;,
\end{equation}
with $\psi$ the electronic operators, $e$ the electronic charge, and $\mathbf{A}(\mathbf{r},t)$ the vector potential.
$\mathbf{A}(\mathbf{r},t)$ includes the external laser with frequency $\omega_L$ and the cavity modes $a_{\perp,\parallel}$ with frequencies $\omega_{\perp,\parallel}$.
By choosing the proper polarization for the cavity modes and the laser, the scheme depicted in Fig.~\ref{fig:fig1}  {(b)} can be realized: the laser and the cavity mode $a_\perp$ induce transitions between the orbital $1$ and $3$, while the cavity mode $a_\parallel$ couples only the orbitals $2$ and $3$.
As we assume the electrons to be localized by the strong interaction between particles, due to the strong localization the field operators can be conveniently expanded onto localized orbitals~\cite{Auerbach_book}: $\psi(\mathbf{r})=\sum_{i,b=1,2,3}w_{ib}(\mathbf{r})c_{ib}$, where $w_{ib}(\mathbf{r})=w_{b}(\mathbf{r}-\mathbf{r}_i)$ with $\mathbf{r}_i$  the position of the center of the unit cell. Here the index $i$ runs over the lattice sites and $b$ is the band index. The interaction Hamiltonian~\eqref{eq:int_ham} thus reads (see Methods for further details)
\begin{equation}
\label{eq:int_ham_sc}
 H_\text{int} \!=\!\sum_i \!\left[c_{i3}^\dag c_{i1}\big( \rho_{31}^L+a_\perp \rho_{31}^\perp\big)+c_{i3}^\dag c_{i2}a_\parallel\rho_{32}^\parallel +\text{h.c.}\right],
\end{equation}
where we neglected counter-rotating terms and changed to the frame rotating with the laser frequency, where $c_{i3}\to c_{i3}e^{-i\omega_L t}$, $a_\ell\to a_\ell e^{-i\omega_L t}$. Correspondingly, the third electron band and the fundamental frequencies of the cavity modes $\omega_\ell$ are shifted as $\Delta_3=\epsilon_3-\omega_L$ and $\Delta_\ell = \omega_\ell-\omega_L$. The matrix elements $\rho_{bb'}^\ell$, $\ell \in \{L,\perp, \parallel\}$ correspond to the transition rates between the bands $b$ and $b'$.

The effective cavity-spin coupling is then obtained by eliminating the third band adiabatically, assuming the band detuning $|\Delta_3|$ to be much larger than the matrix elements $\rho_{bb'}^\ell$, $\ell \in \{L,\perp, \parallel\}$ and the cavity detunings $\Delta_{\perp, \parallel}$. The resulting interaction Hamiltonian describes spins coupled to global, quantum-mechanically fluctuating effective magnetic fields:
\begin{equation}
\label{eq:int_ham_sc_spin_cavity}
 H_{\rm int} = \sum_i \left(  {B}^x  {S}^x_i +  {B}^y  {S}^y_i +  {B}^z  {S}^z_i \right),
\end{equation}
with $ {B}^z = -(\rho_{13}^L\rho_{31}^\perp  {a}_\perp + \text{h.c.})/\Delta_3$, $ {B}^x=-(\rho_{13}^L\rho_{32}^\parallel  {a}_\parallel + \text{h.c.})/\Delta_3$, and $ {B}^y = -(i\rho_{13}^L\rho_{32}^\parallel  {a}_\parallel + \text{h.c.})/\Delta_3$, and $\mathbf{S}_{i}~=~c_{i b}^{\dagger} \boldsymbol{\sigma}_{bb'} c_{i b'} / 2$ is the pseudo-spin operator.
The values of the effective fluctuating effective magnetic fields $ {B}^a$, $a=x,y,z$,  reflect the laser-assisted processes illustrated in Fig.~\ref{fig:fig1}. For instance, $ {B}^x$ and $ {B}^y$, which couple the first and second orbital, result from the laser-assisted excitation of an electron from the first to the third auxiliary band, followed by a decay to the second band with the emission of a cavity photon. The U(1) symmetry of the Hamiltonian results from neglecting the counter-rotating terms, and it is evident from the fact that an excitation from the first to the second band is accompanied only by the creation of a cavity photon, and viceversa. 
Equation~\eqref{eq:int_ham_sc_spin_cavity} is one of the main results of this paper: the effective quantum magnetic fields $ {B}^a$ couple to all the spins, generating an effective long-range coupling.
To further consolidate this insight, we integrate out the cavity field at the level of the Heisenberg equations and obtain an effective Hamiltonian for the spins only~\cite{Piazza2019}: 
\begin{align}
  \label{eq:int_ham_sc_spin}
  H_{\rm int}=&\sum_{ij}\bigg[ \gamma_z S_i^zS_j^z+\gamma_\perp(S_i^xS_j^x+S_i^yS_j^y)\bigg],
\end{align}
with the long-range exchange $\gamma_z = |\rho_{13}^L\rho_{13}^\perp|^2/(\Delta_3^2 \Delta_\perp)$ and $\gamma_\perp = |\rho_{13}^L\rho_{23}^\parallel|^2/(\Delta_3^2 \Delta_\parallel)$.
The interaction is thus naturally U(1)-symmetric, and full SU(2) symmetry can be achieved by adjusting the cavity-mode detunings. Importantly, by choosing the latter to be positive (i.e. a blue-detuned laser), the cavity-mediated interaction is antiferromagnetic.

We now briefly show how multi-mode cavities can generate spatially dependent effective spin-spin interactions.
To this end, we consider a cavity with a large number of modes. For simplicity, we assume them to correspond to photons propagating as plane-waves along the transverse direction with a dispersion  $\Delta_{\ell,\qq} = \sqrt{\omega_\ell^2+(c\,\qq)^2} - \omega_L$, with $c$ the speed of light in the medium.
The form of the Hamiltonian~\eqref{eq:int_ham_sc_spin_cavity} is then preserved, with the fluctuating magnetic fields now possessing a spatial structure according to
\begin{equation}
\label{eq:B-multimode}
 {B}^a = \sum_\qq g^a_\qq \,  {a}_\qq \,\ee^{i\qq\cdot \rr_j} + \text{h.c.},
\end{equation}
with $g^a_\qq $ the momentum-dependent version of the coupling reported below Eq.~\eqref{eq:int_ham_sc_spin_cavity}.
By integrating out the cavity photons, one obtains an effective Hamiltonian as in Eq.~\eqref{eq:int_ham_sc_spin}, where the effective exchange interaction between the spins $ {S}^a$ and $ {S}^b$ is given by
$
\Gamma^{ab}_{ij} ~=~\sum_\qq g^a_\qq g^b_\qq \,\ee^{-i\qq\cdot \rr_{ij}}/\Delta_{\ell,\qq}.
$
While the precise form of $\Gamma^{ab}_{ij}$ depends on the details of $g^a_\qq$, its spatial structure is expected to be long-ranged. In fact, the length scale governing the spatial behaviour is proportional to $ \Delta\ell^{-1/2}$: in THz cavities, the ratio between the lattice size and this lengthscale is of order $10^{-4}$, see, e.g., Ref.~\onlinecite{Schlawin2019}, and therefore $\Gamma^{ab}_{ij}$ can be effectively modelled as a slowly decaying function. For photonic crystal cavities, the form of $\Gamma^{ab}_{ij}$ can be even further engineered by exploiting the band dispersion of the cavity photons~\cite{joannopoulos2008}. The precise form of this function is not expected to qualitatively affect the phase diagram. Accordingly, we choose to parametrize the interaction as $\Gamma^{ab}_{ij} \simeq |\rr_i-\rr_j|^{-\alpha}$, with the value of $\alpha$ compactly encoding the interaction range.  \newtext{The values of $\alpha$ achievable with realistic cavity parameters are of order 10$^{-1}$, and therefore favourable to observe the SL phases (see Methods for further details)}

We finally provide an estimate for the values of $\gamma$ in Eq.~\eqref{eq:Hamiltonian} achievable with this setup (see Methods for further details).
The dipole matrix elements can be estimated assuming a lattice spacing of few angstroms. For THz cavities with a compression factor of $\sim 10^{-5}$ or smaller, a drive with intensity of  $\sim 10$ MW cm$^{-2}$ leads to values of $\gamma$ of order $\sim 100$ K. This number is comparable or larger than typical couplings in antiferromagnets, which range from $\sim 5$ K for vanadates~\cite{Melzi2000} to $\sim 600$ K for iridates~\cite{Kim2012}. For $\alpha$-RuCl$_3$, the (ferromagnetic) Heisenberg interaction is $\sim 40$ K, while the Kitaev one is $\sim 80$ K, see Ref.~\onlinecite{Banerjee2016}. Accordingly, the spin-liquid phases predicted in the phase-diagram in Fig.~\ref{fig:fig1}  {(c)} are achievable with current setups.

%

In this work, we showed that long-range spin-exchange interactions can be robustly induced by coupling a strongly correlated electron system to the quantum fluctuations of a driven cavity.  The electron-cavity coupling gives rise to a variety of tunable spin interactions, including frustrated ones. The thus created cavity-mediated frustration can destroy the magnetic order, favoring disordered spin-liquid states, absent in the cavity-less configuration. We have demonstrated this for an ordinary Heisenberg antiferromagnet, whose ground state manifests an extensive and robust quantum spin-liquid phase when coupled to a cavity. 
Our results open avenues for engineering quantum spin liquids, sparking the challenge to devise new schemes to control electronic degrees of freedom with quantum light, and to uncover phases of matter that are usually inaccessible. This also represents an exciting perspective for the experimental detection of strongly correlated phases: photons emitted from the cavities carry signatures of the quantum many-body state, which become accessible to standard optical measurements. 
Our findings are immediately relevant also for quantum simulations. Artificial spin systems with tunable long-range interactions can be currently created using either trapped ions~\cite{jurcevic2014,richerme2014} or ultracold atoms coupled to an optical cavity~\cite{Landini2018,Kroeze2018,Krevsic2018,Davis2019}. These platforms represents therefore ideal candidates to simulate quantum spin liquid phases.

\noindent \textbf{Acknowledgments.} We warmly acknowledge discussions with C.~Hickey, G.~Mazza, A.~Rosch, M.~Scherer, and especially S.~Trebst. We acknowledge support by the funding from the European Research Council (ERC) under the Horizon 2020 research and innovation program, Grant Agreement No. 647434 (DOQS), and by the DFG Collaborative Research Center (CRC) 1238 Project No. 277146847 - projects C02, C03, and C04.

\noindent \textbf{Author contributions.} S.~D. and A.~C. designed the research, D.~K. performed the numerical simulations, A.~C. and F.~P. developed the implementation scheme. \newtext{C.~P.~Z. computed and analysed the dynamical structure factor data.} All authors analyzed the results and contributed to the manuscript.

\noindent \textbf{Competing interests.} The authors declare no competing interests. 

\noindent \textbf{Data and material availability.}  The code that supports the plots
within this paper are available from the corresponding author upon reasonable request.

\section*{Methods}

\subsection*{Saddle-point equations for bosonic spinons}
%
In this Section we outline the derivation of the saddle point equations for the spinon bilinear expectation values $\A_{ij}$ and $\B_{ij}$.

The spin exchange terms appearing in Eq.~\eqref{eq:Hamiltonian} can be recast as $\SP_i\SP_j = :\! B^\dagger_{ij} B_{ij} \!: - A^\dagger_{ij}A_{ij}$ for $i\neq j$, where $A_{ij} = i \sigma^y_{\mu\nu} b_{i\mu}b_{j\nu}/2,$ and $B_{ij} =  b^\dagger_{i\mu}b_{j\mu}/2$ are SU(2)-invariant spinonic bilinears. A finite expectation value of these operators indicates the tendency of the spins at the sites $i$ and $j$ of forming a singlet ($A_{ij}$) or to align ($B_{ij}$): moreover, it induces a finite bosonic hopping rate between the lattice sites $i$ and $j$, signalling the existence of propagating fractional excitations. 
In order to solve for the value of these quantities, we build on the approach of Ref.~\cite{Arovas1988}. First, the bosonized version of the Hamiltonian~\eqref{eq:Hamiltonian} is represented as a path integral, with the constraint implemented by a space- and time-dependent Lagrange multiplier $\lambda_i(t)$. After decoupling the bilinear products by using a Hubbard-Stratonovich transformation, the expectation values $\A_{ij} = \langle A_{ij} \rangle$ and $\B_{ij} = \langle B_{ij}\rangle$ are  obtained as saddle point values of the corresponding action. This approximation imposes the constraint only on average, and the now position- and time-independent Lagrange multiplier $\lambda$ has to be determined self-consistently.
This approximation is equivalent to decoupling the Hamiltonian~\eqref{eq:Hamiltonian} in bosonic bilinears as:
\begin{equation}
\label{eq:HF-Hamiltonian}
H 
= \frac{1}{2}\sum_{ i,j}\left( \epsilon_{ij}  b^\dagger_{i\mu}b_{j\mu}
+i\Delta^*_{ij} \sigma^y_{\mu\nu}  b_{i\mu}b_{j\nu} \right)
+ \text{h.c.} + \varepsilon_0,
\end{equation}
where $\epsilon_{ij} = J_{ij} \B^*_{ij} + \delta_{ij} \lambda/2$,  $\Delta^*_{ij} = - J_{ij}\A^*_{ij}$, and $\varepsilon_0 = \sum_{ i,j} (- |\B_{ij}|^2 + |\A_{ij}|^2) -2 S N \lambda $.
As discussed in the main text, we assume a translational-invariant ansatz, i.e., $\A_{ij} =\A_{i-j}$ and $\B_{ij} =\B_{i-j}$, 
able to interpolate between all the expected phases. The two degenerate eigenvalues of $H$ are  given by $E^2_\qq = \epsilon^2_\qq -|\Delta_\qq|^2$, with $\epsilon_\qq$ and $\Delta_\qq$ the Fourier transform of the functions appearing in Eq.~\eqref{eq:HF-Hamiltonian}. By minimizing the ground-state energy $E_0 = \sum_\qq\ \left(E_\qq -\epsilon_\qq \right) + \varepsilon_0$ with respect to the variational parameters $\epsilon_\qq$, $\Delta_\qq$, and $\lambda$, one obtains the saddle-point equations:
\begin{subequations}
\label{eq:HF-equations}
\begin{align}
&1 = \frac{1}{2N} \sum_\qq \frac{\epsilon_\qq}{E_\qq}, \\
&\epsilon_\pp  = \lambda + \frac{1}{2N}\sum_{\qq} J_{\pp-\qq}\left(\frac{\epsilon_\qq}{E_\qq}-1\right), \\
&\Delta_\pp  =  \frac{1}{2N}\sum_{\qq} J_{\pp-\qq} \frac{\Delta_\qq}{E_\qq},
\end{align}
\end{subequations}
with $J_\qq$ the Fourier transform of $J_{ij}$. These equations provide the full momentum dependence of the functions $\epsilon_\qq$ and $\Delta_\qq.$ The actual number of unknowns increases with the range of the interaction. In fact, for short-range interactions the momentum dependence can be found analytically, and only few parameters are left to be computed self-consistently. For long-range interactions, instead, the full momentum dependence needs to be found numerically. Eqs.~\eqref{eq:HF-equations}  amount to a system of $2N + 1$ coupled non-linear equations, with $N$ the total number of sites. {To find the roots of these equations, we used a trust region solver as provided by the Julia NLSolve library, with the accepted residual norm set to $10^{-8}$. The error of the numerical solution for a finite system of $N$ sites is therefore negligible compared to the extrapolation to the thermodynamic limit. Determining $\xi_{\mathcal{O}}(L)$ by a least-squares fit resulted in relative errors between $0.5\% - 2 \%$, which we consider to be sufficient for the analysis performed here. The root finding algorithm is accelerated by exploiting vectorization for the evaluation of the saddle-point equations where possible, and by parallelization via OpenBLAS.}

\subsection*{Implementation details}
%
Here we provide additional details to the setup described in the main text.
The vector potential can be written as
\begin{equation}
 {\mathbf{A}}(\mathbf{r}, t)=\Omega_{L} \mathbf{u}_{L} \varphi_{L}(\mathbf{r}) e^{i \omega_{L} t}+\sum_{\ell=\|, \perp} \mathcal{N}_{\ell} \mathbf{u}_{\ell}  {a}_{\ell} \varphi_{\ell}(\mathbf{r})+\mathrm{h.c.},
\end{equation}  
where $\Omega_{L}^2$ and $\omega_L$ denote the laser intensity and frequency, respectively. Here $\mathbf{u}$ and $\varphi(\mathbf{r})$ are the polarization vector and the mode wavefunction. For the cavity modes, labelled by $\ell=\|, \perp,$ the wave function is normalized over the finite volume $V_{c}$ and $\mathcal{N}_{\ell}=\sqrt{1 / 2 \omega_{\ell} \epsilon_{0} \epsilon_{r}},$ where $\omega_{\ell}$ is the mode fundamental frequency and $\epsilon_{0}, \epsilon_{r}$ are the vacuum and relative permittivity of the material, respectively. 

We assume that that the mode wavefunctions $\varphi(\mathbf{r})$ does not vary significantly over the extent of the Wannier functions.
By tuning the polarization vectors $\mathbf{u}$ to selectively couple the orbitals as in Fig.~\ref{fig:fig1}, and by performing the rotating-wave approximation, the resulting paramagnetic Hamiltonian term is given by Eq.~\eqref{eq:int_ham_sc}, with
\begin{equation}
\rho_{b b^{\prime}}^{\ell} =\frac{e \mathcal{N}_{\ell}}{m} \varphi_\ell \mathbf{u}_{\ell}\cdot \langle w_{i b} |\pp | w_{ib'}\rangle. 
\end{equation}
The expression for $\rho_{b b^{\prime}}^{L}$ can be obtained from the previous equation
by replacing $\mathcal{N}_\ell$ with $\Omega_L$.

The diamagnetic part of the Hamiltonian~\eqref{eq:int_ham} reads, after neglecting higher-order electron-photon processes of the type $ {c}^\dagger_3 {c}_3( {a}+ {a}^\dagger)$ and $ {c}^\dagger_3 {c}_3 {a}^\dagger {a}$:
\begin{equation}
\label{eq:RWA-dia}
 {H}_\text{int,dia} = \sum_{\ell=\perp,\parallel} \delta_\ell\,   {a}^\dagger_\ell {a_\ell} +\sum_{i} \delta_3\,  {c}^\dagger_{i3} {c}_{i3},
\end{equation}
plus a term linear in the cavity fields, which vanishes as the laser and cavity wavefunctions are orthogonal.
The shifts
\begin{subequations}
\begin{align}
\delta_\ell &=  \frac{e^2}{m} V_\text{e}\,  \mathcal{N}_\ell^2 |\varphi_\ell|^2, \\
\delta_3 & =
\frac{e^2 \Omega_L^2 }{m} |\varphi_L|^2  + \sum_{\ell=\perp,\parallel} \frac{e^2\mathcal{N}_\ell^2}{2m} |\varphi_\ell|^2,
\end{align}
\end{subequations}
\begin{figure}[t!]
\includegraphics[width=8cm]{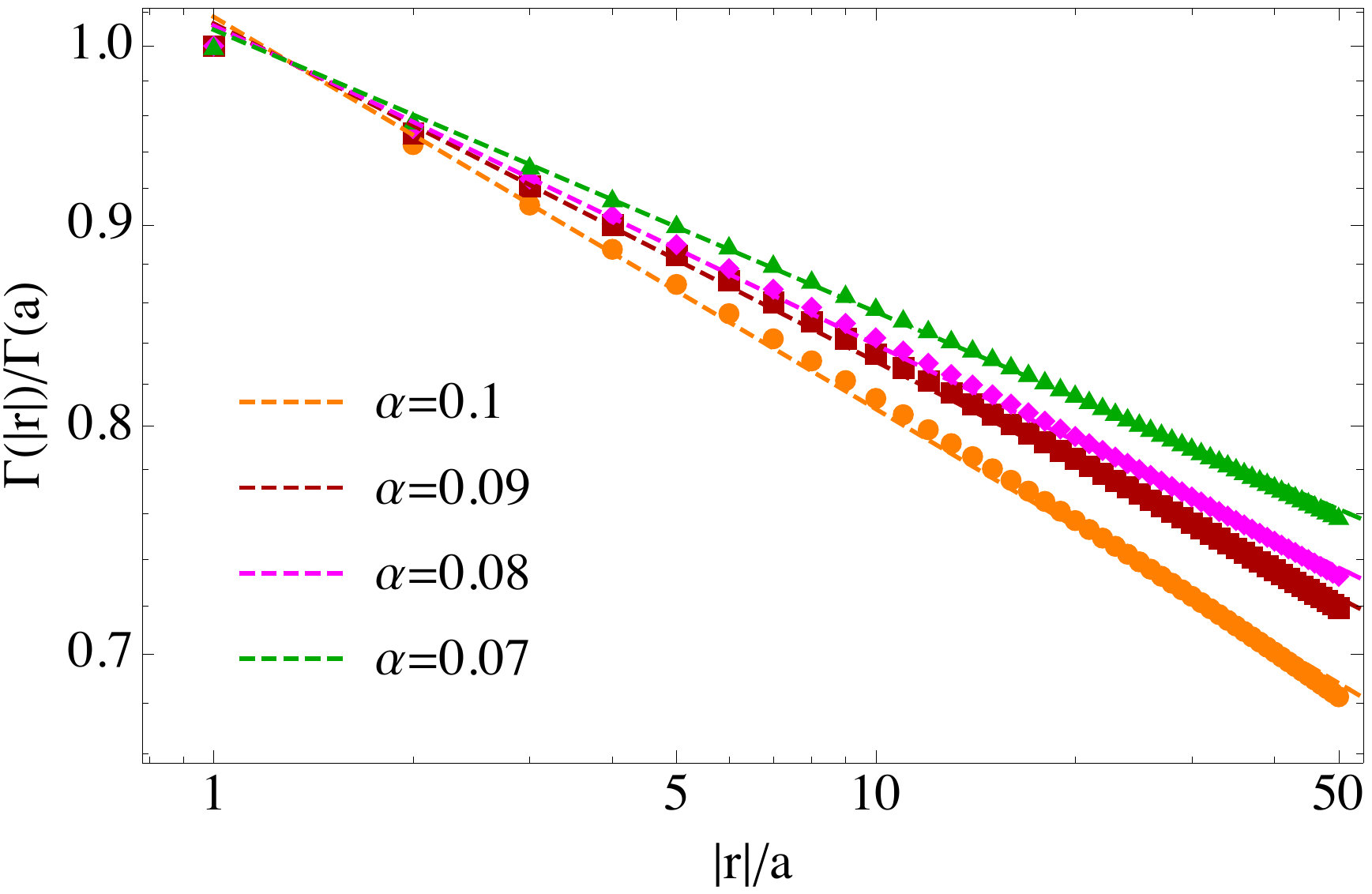}
\caption{Normalized long-range spin exchange $\Gamma(|\rr|)$ as a function of the distance between sites $|\rr|$. The cavity fundamental wavelength is $\lambda =  2\pi c/\omega = 10^5 a$, with $a$ the lattice spacing. The symbols correspond to different values of the cavity detuning: $\Delta = 0.5 \omega$ (orange), $\Delta = 0.1\omega$ (red), $\Delta = 0.05\omega$ (purple), and $\Delta = 0.01 \omega$ (green). The dashed curves correspond to the fitted power-laws, with exponents reported in the legend. 
} 
\label{fig:fig3}
\end{figure}
renormalize the energies of the cavity modes and of the third band, respectively.  
By assuming that the band detuning $|\widetilde{\Delta}_3|$ is much larger than the coupling strengths and the cavity detunings $\widetilde{\Delta}_\ell$, the third band can be adiabatically eliminated, leading to Eq.~\eqref{eq:int_ham_sc_spin_cavity} in the main text, including an additional term $ B_0 \sum_{i} {S}^z_i$, with $B_0  = |\rho_{13}^L|^2/\widetilde{\Delta}_3$. This effective classical magnetic field breaks explicitly the SU(2) symmetry, but it is much smaller than the spin exchange and therefore it can be safely neglected.

\subsection*{\newtext{Estimate of interaction strength and range}}
%
We consider a THz laser ($\omega_L = 100$ THz) with intensity $\Omega_L^2 = 10$ MW cm$^{-2}$, with a small detuning from the cavity frequency $\Delta_\perp = \Delta_\parallel = 10^{-2}$ THz. The compression factor of the cavity is assumed to be $\Lambda = 10^{-5}$. The detuning from the third band is $\Delta_3 = 1$ THz, thus satisfying the condition $\Delta_3 \gg \Delta_\perp$. We estimate the matrix as follows: $\langle w_{i 1} |\pp | w_{i3'}\rangle \sim m \omega_{13} \langle w_{i 1} |\rr | w_{i3'}\rangle $, with $\omega_{13} = \omega_L +\Delta_L$ and $\langle w_{i 1} |\rr | w_{i3'}\rangle = 10 \AA$, the same order of magnitude of a typical lattice spacing. Using the formulas derived in the text, one then estimates a long-range interaction with strength $\gamma \sim 100$ K.

We also provide an estimate of the values of $\alpha$. To this end, we evaluate the explicit form of $\Gamma(\rr_{ij})$ as reported in the text below Eq.~\eqref{eq:B-multimode}. \newtext{For simplicity, we assume $\Delta_\perp = \Delta_\parallel \equiv \Delta$ and $\omega_\perp = \omega_\parallel \equiv \omega$. The Rabi-like couplings $g_\qq$ inherits the momentum dependence from the normalization of every mode, i.e., $g_\qq \propto (\omega^2 +(c\qq)^2)^{-1/4}$.  Accordingly, the cavity-mediated exchange is given by:  
\begin{equation}
  \Gamma(\rr_{ij}) \propto  \sum_\qq  \frac{\ee^{-i \qq\cdot \rr_{ij}}}{\sqrt{\omega^2 + (c\qq)^2}(\sqrt{\omega_c^2 + (c\qq)^2}-\omega_L)}.
\end{equation}
}
The corresponding integral is computed numerically, and the results shown in Fig.~\ref{fig:fig3} over a range of 50 lattice sites, for different values of the cavity detuning.  The values of $\alpha$ obtained are reported in the figure.

\begin{figure}
    \centering
    \includegraphics[width=8cm]{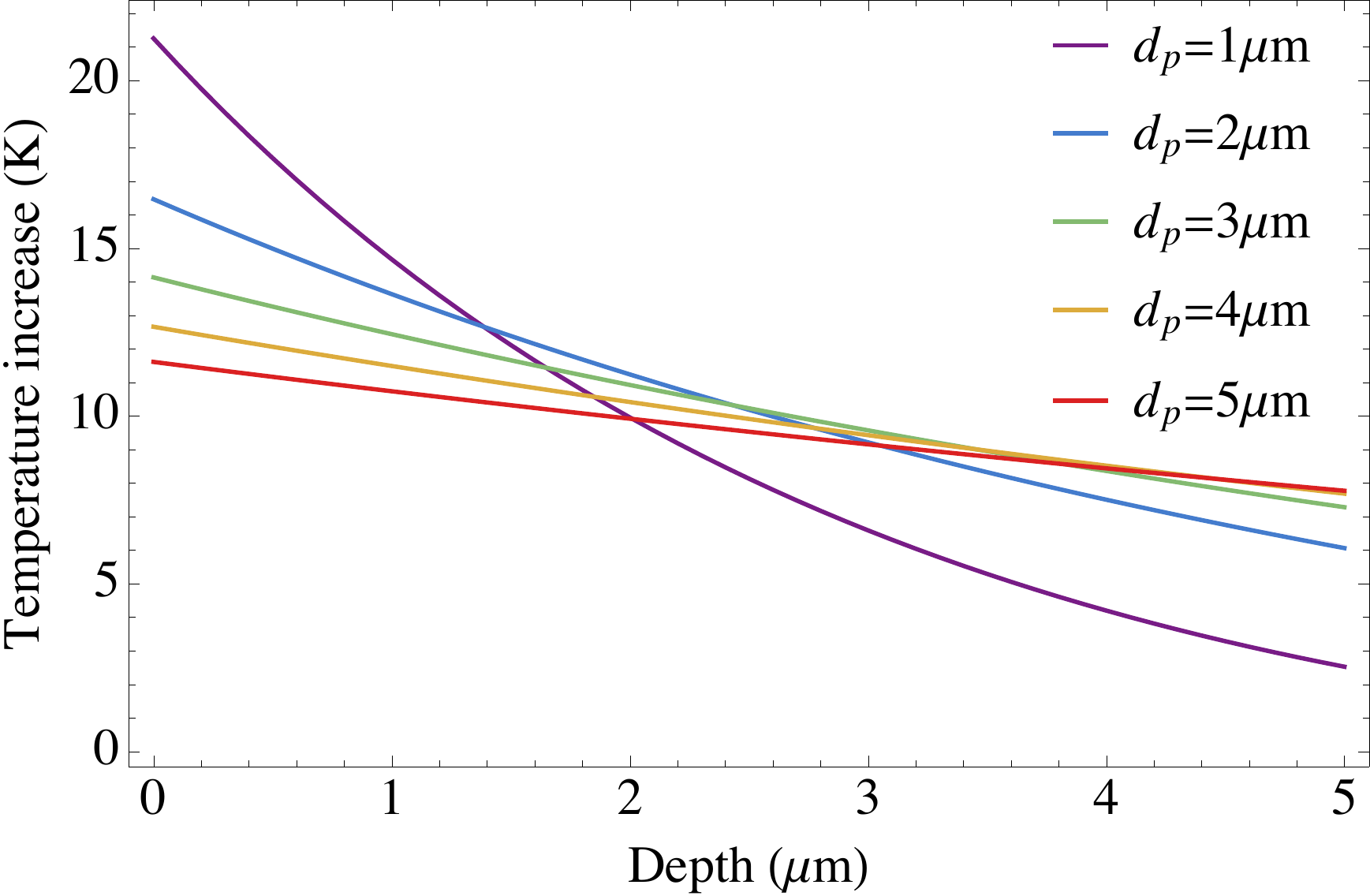}
    \caption{\newtext{Laser-induced temperature increase as a function of the depth in the sample, for different values of the laser penetration depths.}}
    \label{fig:heating}
\end{figure}

\subsection*{\newtext{Heating effects}}

\newtext{
A possible advantage of our scheme is that it does not rely on the laser being resonant with any electronic or phononic excitation, with the only tradeoff of a decreasing coupling strength as the detuning increases. Exploiting the variability of the detuning, as well as the knowledge of the relevant excitation-modes of the quantum material and the cavity, resonances can be avoided and heating be pushed to later times. 
This said, we can on the other hand consider the worst-case scenario, where heating does efficiently take place, and estimate the amount of it, following Ref.~\onlinecite{Gao}. As a paradigmatic material, we consider RuCl$_3$, which features orbital pseudospin, and whose low-temperature properties have been intensely studied~\cite{Banerjee2016}. In order to estimate the heating of the material surface due to the external laser, we evaluate the energy density deposited by the laser using the formula:
\begin{equation}
    \epsilon(z) = (1-R) \frac{\mathcal{F}}{d_{\text {p }}} \mathrm{e}^{-\frac{z}{d_\text{p}}},
\end{equation}
with $z$ the depth in the sample, $R\sim 0.05$ the material reflectivity at the laser frequency considered, i.e., 100 THz (cf. Ref.~\onlinecite{Loidl}), $d_\text{p}$ the penetration depth, chosen of order of micrometers, and $\mathcal{F}$ is the excitation energy density. In order to achieve lasing in the desired frequency range, a short-pulse protocol can be used. By considering pulses of $\sim$ 10 ps, and a maximal laser intensity of 10 MW/cm$^2$, we require $\mathcal{F} =  10^{-5}$ J/cm$^2$. 
In order to estimate the increase in temperature due to the deposited energy density $\epsilon(z)$, we assume that thermalization time is fast, and use the following relation:
\begin{equation}
    \epsilon(z) = \frac{1}{V_m}\int_{T_0}^{T_f(z)} C_p(T) \mathrm{d} T, 
\end{equation}
where $V_m = 53.32 \text{cm}^3/\text{mol}$ is the molar volume of $\alpha-$RuCl$_3$, and  $C_p$ is the molar heat capacity, for which we use the value fitted from the measurements of Ref.~\onlinecite{Cao}, and $T_0$ is the initial temperature in the sample. For an initial temperature of $T_0=2K$, the rise in temperature as a function of $z$ and for different values of the penetration length $d_\text{p}$ are reported in Fig.~\ref{fig:heating} here.
The estimated temperature increase in the first layers is is in between 5 K and 20 K, depending on the penetration depth. In order to understand the impact of heating on the expected QSL phases, a simple criterion compares the temperature increase with the gap (at least for the gapped spin liquid phase, which we dubbed SL-I in the main text). Robustness of the QSL phase then requires the temperature increase to be smaller than the gap, whose scale is given by the material couplings and by the cavity-induced interaction. For $\alpha-$RuCl$_3$, as discussed in the main text, interactions lie between 40K and 80K. Accordingly, the heating induced by the laser is not expected to destabilize the gapped QSL phase. For what concerns the gapless QSL phase (SL-II in the main text), the previous argument is clearly inapplicable, and its robustness against thermal fluctuations must be assessed using a more sophisticated approach, e.g., by solving the saddle-point equations at finite temperature.
}

\bibliography{cQSL_biblio}

\begin{thebibliography}{54}%
\makeatletter
\providecommand \@ifxundefined [1]{%
 \@ifx{#1\undefined}
}%
\providecommand \@ifnum [1]{%
 \ifnum #1\expandafter \@firstoftwo
 \else \expandafter \@secondoftwo
 \fi
}%
\providecommand \@ifx [1]{%
 \ifx #1\expandafter \@firstoftwo
 \else \expandafter \@secondoftwo
 \fi
}%
\providecommand \natexlab [1]{#1}%
\providecommand \enquote  [1]{``#1''}%
\providecommand \bibnamefont  [1]{#1}%
\providecommand \bibfnamefont [1]{#1}%
\providecommand \citenamefont [1]{#1}%
\providecommand \href@noop [0]{\@secondoftwo}%
\providecommand \href [0]{\begingroup \@sanitize@url \@href}%
\providecommand \@href[1]{\@@startlink{#1}\@@href}%
\providecommand \@@href[1]{\endgroup#1\@@endlink}%
\providecommand \@sanitize@url [0]{\catcode `\\12\catcode `\$12\catcode
  `\&12\catcode `\#12\catcode `\^12\catcode `\_12\catcode `\%12\relax}%
\providecommand \@@startlink[1]{}%
\providecommand \@@endlink[0]{}%
\providecommand \url  [0]{\begingroup\@sanitize@url \@url }%
\providecommand \@url [1]{\endgroup\@href {#1}{\urlprefix }}%
\providecommand \urlprefix  [0]{URL }%
\providecommand \Eprint [0]{\href }%
\providecommand \doibase [0]{http://dx.doi.org/}%
\providecommand \selectlanguage [0]{\@gobble}%
\providecommand \bibinfo  [0]{\@secondoftwo}%
\providecommand \bibfield  [0]{\@secondoftwo}%
\providecommand \translation [1]{[#1]}%
\providecommand \BibitemOpen [0]{}%
\providecommand \bibitemStop [0]{}%
\providecommand \bibitemNoStop [0]{.\EOS\space}%
\providecommand \EOS [0]{\spacefactor3000\relax}%
\providecommand \BibitemShut  [1]{\csname bibitem#1\endcsname}%
\let\auto@bib@innerbib\@empty
\bibitem [{\citenamefont {Savary}\ and\ \citenamefont
  {Balents}(2016)}]{Balents_review}%
  \BibitemOpen
  \bibfield  {author} {\bibinfo {author} {\bibfnamefont {L.}~\bibnamefont
  {Savary}}\ and\ \bibinfo {author} {\bibfnamefont {L.}~\bibnamefont
  {Balents}},\ }\href {\doibase 10.1088/0034-4885/80/1/016502} {\bibfield
  {journal} {\bibinfo  {journal} {Reports on Progress in Physics}\ }\textbf
  {\bibinfo {volume} {80}},\ \bibinfo {pages} {016502} (\bibinfo {year}
  {2016})}\BibitemShut {NoStop}%
\bibitem [{\citenamefont {Zhou}\ \emph {et~al.}(2017)\citenamefont {Zhou},
  \citenamefont {Kanoda},\ and\ \citenamefont {Ng}}]{Ng_review}%
  \BibitemOpen
  \bibfield  {author} {\bibinfo {author} {\bibfnamefont {Y.}~\bibnamefont
  {Zhou}}, \bibinfo {author} {\bibfnamefont {K.}~\bibnamefont {Kanoda}}, \ and\
  \bibinfo {author} {\bibfnamefont {T.-K.}\ \bibnamefont {Ng}},\ }\href
  {\doibase 10.1103/RevModPhys.89.025003} {\bibfield  {journal} {\bibinfo
  {journal} {Rev. Mod. Phys.}\ }\textbf {\bibinfo {volume} {89}},\ \bibinfo
  {pages} {025003} (\bibinfo {year} {2017})}\BibitemShut {NoStop}%
\bibitem [{\citenamefont {Knolle}\ and\ \citenamefont
  {Moessner}(2019)}]{Moessner_review}%
  \BibitemOpen
  \bibfield  {author} {\bibinfo {author} {\bibfnamefont {J.}~\bibnamefont
  {Knolle}}\ and\ \bibinfo {author} {\bibfnamefont {R.}~\bibnamefont
  {Moessner}},\ }\href {\doibase 10.1146/annurev-conmatphys-031218-013401}
  {\bibfield  {journal} {\bibinfo  {journal} {Annual Review of Condensed Matter
  Physics}\ }\textbf {\bibinfo {volume} {10}},\ \bibinfo {pages} {451}
  (\bibinfo {year} {2019})}\BibitemShut {NoStop}%
\bibitem [{\citenamefont {Balents}(2010)}]{Balents_frustration}%
  \BibitemOpen
  \bibfield  {author} {\bibinfo {author} {\bibfnamefont {L.}~\bibnamefont
  {Balents}},\ }\href@noop {} {\bibfield  {journal} {\bibinfo  {journal}
  {Nature}\ }\textbf {\bibinfo {volume} {464}},\ \bibinfo {pages} {199}
  (\bibinfo {year} {2010})}\BibitemShut {NoStop}%
\bibitem [{\citenamefont {Castelnovo}\ \emph {et~al.}(2012)\citenamefont
  {Castelnovo}, \citenamefont {Moessner},\ and\ \citenamefont
  {Sondhi}}]{Castelnovo2012}%
  \BibitemOpen
  \bibfield  {author} {\bibinfo {author} {\bibfnamefont {C.}~\bibnamefont
  {Castelnovo}}, \bibinfo {author} {\bibfnamefont {R.}~\bibnamefont
  {Moessner}}, \ and\ \bibinfo {author} {\bibfnamefont {S.}~\bibnamefont
  {Sondhi}},\ }\href {\doibase 10.1146/annurev-conmatphys-020911-125058}
  {\bibfield  {journal} {\bibinfo  {journal} {Annual Review of Condensed Matter
  Physics}\ }\textbf {\bibinfo {volume} {3}},\ \bibinfo {pages} {35} (\bibinfo
  {year} {2012})}\BibitemShut {NoStop}%
\bibitem [{\citenamefont {Gingras}\ and\ \citenamefont
  {McClarty}(2014)}]{Gingras2014}%
  \BibitemOpen
  \bibfield  {author} {\bibinfo {author} {\bibfnamefont {M.~J.~P.}\
  \bibnamefont {Gingras}}\ and\ \bibinfo {author} {\bibfnamefont {P.~A.}\
  \bibnamefont {McClarty}},\ }\href {\doibase 10.1088/0034-4885/77/5/056501}
  {\bibfield  {journal} {\bibinfo  {journal} {Reports on Progress in Physics}\
  }\textbf {\bibinfo {volume} {77}},\ \bibinfo {pages} {056501} (\bibinfo
  {year} {2014})}\BibitemShut {NoStop}%
\bibitem [{\citenamefont {Norman}(2016)}]{Norman2016}%
  \BibitemOpen
  \bibfield  {author} {\bibinfo {author} {\bibfnamefont {M.~R.}\ \bibnamefont
  {Norman}},\ }\href {\doibase 10.1103/RevModPhys.88.041002} {\bibfield
  {journal} {\bibinfo  {journal} {Rev. Mod. Phys.}\ }\textbf {\bibinfo {volume}
  {88}},\ \bibinfo {pages} {041002} (\bibinfo {year} {2016})}\BibitemShut
  {NoStop}%
\bibitem [{\citenamefont {Yao}\ \emph {et~al.}(2018)\citenamefont {Yao},
  \citenamefont {Zaletel}, \citenamefont {Stamper-Kurn},\ and\ \citenamefont
  {Vishwanath}}]{Yao2018}%
  \BibitemOpen
  \bibfield  {author} {\bibinfo {author} {\bibfnamefont {N.~Y.}\ \bibnamefont
  {Yao}}, \bibinfo {author} {\bibfnamefont {M.~P.}\ \bibnamefont {Zaletel}},
  \bibinfo {author} {\bibfnamefont {D.~M.}\ \bibnamefont {Stamper-Kurn}}, \
  and\ \bibinfo {author} {\bibfnamefont {A.}~\bibnamefont {Vishwanath}},\
  }\href@noop {} {\bibfield  {journal} {\bibinfo  {journal} {Nature Physics}\
  }\textbf {\bibinfo {volume} {14}},\ \bibinfo {pages} {405} (\bibinfo {year}
  {2018})}\BibitemShut {NoStop}%
\bibitem [{\citenamefont {Zou}\ \emph {et~al.}(2017)\citenamefont {Zou},
  \citenamefont {Zhao},\ and\ \citenamefont {Liu}}]{Zou2017}%
  \BibitemOpen
  \bibfield  {author} {\bibinfo {author} {\bibfnamefont {H.}~\bibnamefont
  {Zou}}, \bibinfo {author} {\bibfnamefont {E.}~\bibnamefont {Zhao}}, \ and\
  \bibinfo {author} {\bibfnamefont {W.~V.}\ \bibnamefont {Liu}},\ }\href
  {\doibase 10.1103/PhysRevLett.119.050401} {\bibfield  {journal} {\bibinfo
  {journal} {Phys. Rev. Lett.}\ }\textbf {\bibinfo {volume} {119}},\ \bibinfo
  {pages} {050401} (\bibinfo {year} {2017})}\BibitemShut {NoStop}%
\bibitem [{\citenamefont {Kele\ifmmode~\mbox{\c{s}}\else \c{s}\fi{}}\ and\
  \citenamefont {Zhao}(2018)}]{Keles2018}%
  \BibitemOpen
  \bibfield  {author} {\bibinfo {author} {\bibfnamefont {A.}~\bibnamefont
  {Kele\ifmmode~\mbox{\c{s}}\else \c{s}\fi{}}}\ and\ \bibinfo {author}
  {\bibfnamefont {E.}~\bibnamefont {Zhao}},\ }\href {\doibase
  10.1103/PhysRevLett.120.187202} {\bibfield  {journal} {\bibinfo  {journal}
  {Phys. Rev. Lett.}\ }\textbf {\bibinfo {volume} {120}},\ \bibinfo {pages}
  {187202} (\bibinfo {year} {2018})}\BibitemShut {NoStop}%
\bibitem [{\citenamefont {Sentef}\ \emph {et~al.}(2018)\citenamefont {Sentef},
  \citenamefont {Ruggenthaler},\ and\ \citenamefont {Rubio}}]{Sentef2018}%
  \BibitemOpen
  \bibfield  {author} {\bibinfo {author} {\bibfnamefont {M.~A.}\ \bibnamefont
  {Sentef}}, \bibinfo {author} {\bibfnamefont {M.}~\bibnamefont
  {Ruggenthaler}}, \ and\ \bibinfo {author} {\bibfnamefont {A.}~\bibnamefont
  {Rubio}},\ }\href@noop {} {\bibfield  {journal} {\bibinfo  {journal} {Science
  advances}\ }\textbf {\bibinfo {volume} {4}},\ \bibinfo {pages} {eaau6969}
  (\bibinfo {year} {2018})}\BibitemShut {NoStop}%
\bibitem [{\citenamefont {Mazza}\ and\ \citenamefont
  {Georges}(2019)}]{Mazza2019}%
  \BibitemOpen
  \bibfield  {author} {\bibinfo {author} {\bibfnamefont {G.}~\bibnamefont
  {Mazza}}\ and\ \bibinfo {author} {\bibfnamefont {A.}~\bibnamefont
  {Georges}},\ }\href {\doibase 10.1103/PhysRevLett.122.017401} {\bibfield
  {journal} {\bibinfo  {journal} {Phys. Rev. Lett.}\ }\textbf {\bibinfo
  {volume} {122}},\ \bibinfo {pages} {017401} (\bibinfo {year}
  {2019})}\BibitemShut {NoStop}%
\bibitem [{\citenamefont {Andolina}\ \emph {et~al.}(2019)\citenamefont
  {Andolina}, \citenamefont {Pellegrino}, \citenamefont {Giovannetti},
  \citenamefont {MacDonald},\ and\ \citenamefont {Polini}}]{Andolina2019}%
  \BibitemOpen
  \bibfield  {author} {\bibinfo {author} {\bibfnamefont {G.~M.}\ \bibnamefont
  {Andolina}}, \bibinfo {author} {\bibfnamefont {F.~M.~D.}\ \bibnamefont
  {Pellegrino}}, \bibinfo {author} {\bibfnamefont {V.}~\bibnamefont
  {Giovannetti}}, \bibinfo {author} {\bibfnamefont {A.~H.}\ \bibnamefont
  {MacDonald}}, \ and\ \bibinfo {author} {\bibfnamefont {M.}~\bibnamefont
  {Polini}},\ }\href {\doibase 10.1103/PhysRevB.100.121109} {\bibfield
  {journal} {\bibinfo  {journal} {Phys. Rev. B}\ }\textbf {\bibinfo {volume}
  {100}},\ \bibinfo {pages} {121109} (\bibinfo {year} {2019})}\BibitemShut
  {NoStop}%
\bibitem [{\citenamefont {Schlawin}\ \emph {et~al.}(2019)\citenamefont
  {Schlawin}, \citenamefont {Cavalleri},\ and\ \citenamefont
  {Jaksch}}]{Schlawin2019}%
  \BibitemOpen
  \bibfield  {author} {\bibinfo {author} {\bibfnamefont {F.}~\bibnamefont
  {Schlawin}}, \bibinfo {author} {\bibfnamefont {A.}~\bibnamefont {Cavalleri}},
  \ and\ \bibinfo {author} {\bibfnamefont {D.}~\bibnamefont {Jaksch}},\ }\href
  {\doibase 10.1103/PhysRevLett.122.133602} {\bibfield  {journal} {\bibinfo
  {journal} {Phys. Rev. Lett.}\ }\textbf {\bibinfo {volume} {122}},\ \bibinfo
  {pages} {133602} (\bibinfo {year} {2019})}\BibitemShut {NoStop}%
\bibitem [{\citenamefont {Curtis}\ \emph {et~al.}(2019)\citenamefont {Curtis},
  \citenamefont {Raines}, \citenamefont {Allocca}, \citenamefont {Hafezi},\
  and\ \citenamefont {Galitski}}]{Curtis2019}%
  \BibitemOpen
  \bibfield  {author} {\bibinfo {author} {\bibfnamefont {J.~B.}\ \bibnamefont
  {Curtis}}, \bibinfo {author} {\bibfnamefont {Z.~M.}\ \bibnamefont {Raines}},
  \bibinfo {author} {\bibfnamefont {A.~A.}\ \bibnamefont {Allocca}}, \bibinfo
  {author} {\bibfnamefont {M.}~\bibnamefont {Hafezi}}, \ and\ \bibinfo {author}
  {\bibfnamefont {V.~M.}\ \bibnamefont {Galitski}},\ }\href {\doibase
  10.1103/PhysRevLett.122.167002} {\bibfield  {journal} {\bibinfo  {journal}
  {Phys. Rev. Lett.}\ }\textbf {\bibinfo {volume} {122}},\ \bibinfo {pages}
  {167002} (\bibinfo {year} {2019})}\BibitemShut {NoStop}%
\bibitem [{\citenamefont {Thomas}\ \emph {et~al.}(2019)\citenamefont {Thomas},
  \citenamefont {Devaux}, \citenamefont {Nagarajan}, \citenamefont {Chervy},
  \citenamefont {Seidel}, \citenamefont {Hagenm{\"u}ller}, \citenamefont
  {Sch{\"u}tz}, \citenamefont {Schachenmayer}, \citenamefont {Genet},
  \citenamefont {Pupillo} \emph {et~al.}}]{Thomas2019}%
  \BibitemOpen
  \bibfield  {author} {\bibinfo {author} {\bibfnamefont {A.}~\bibnamefont
  {Thomas}}, \bibinfo {author} {\bibfnamefont {E.}~\bibnamefont {Devaux}},
  \bibinfo {author} {\bibfnamefont {K.}~\bibnamefont {Nagarajan}}, \bibinfo
  {author} {\bibfnamefont {T.}~\bibnamefont {Chervy}}, \bibinfo {author}
  {\bibfnamefont {M.}~\bibnamefont {Seidel}}, \bibinfo {author} {\bibfnamefont
  {D.}~\bibnamefont {Hagenm{\"u}ller}}, \bibinfo {author} {\bibfnamefont
  {S.}~\bibnamefont {Sch{\"u}tz}}, \bibinfo {author} {\bibfnamefont
  {J.}~\bibnamefont {Schachenmayer}}, \bibinfo {author} {\bibfnamefont
  {C.}~\bibnamefont {Genet}}, \bibinfo {author} {\bibfnamefont
  {G.}~\bibnamefont {Pupillo}},  \emph {et~al.},\ }\href@noop {} {\bibfield
  {journal} {\bibinfo  {journal} {arXiv preprint arXiv:1911.01459}\ } (\bibinfo
  {year} {2019})}\BibitemShut {NoStop}%
\bibitem [{\citenamefont {Gao}\ \emph {et~al.}(2020{\natexlab{a}})\citenamefont
  {Gao}, \citenamefont {Schlawin}, \citenamefont {Buzzi}, \citenamefont
  {Cavalleri},\ and\ \citenamefont {Jaksch}}]{Gao2020}%
  \BibitemOpen
  \bibfield  {author} {\bibinfo {author} {\bibfnamefont {H.}~\bibnamefont
  {Gao}}, \bibinfo {author} {\bibfnamefont {F.}~\bibnamefont {Schlawin}},
  \bibinfo {author} {\bibfnamefont {M.}~\bibnamefont {Buzzi}}, \bibinfo
  {author} {\bibfnamefont {A.}~\bibnamefont {Cavalleri}}, \ and\ \bibinfo
  {author} {\bibfnamefont {D.}~\bibnamefont {Jaksch}},\ }\href {\doibase
  10.1103/PhysRevLett.125.053602} {\bibfield  {journal} {\bibinfo  {journal}
  {Phys. Rev. Lett.}\ }\textbf {\bibinfo {volume} {125}},\ \bibinfo {pages}
  {053602} (\bibinfo {year} {2020}{\natexlab{a}})}\BibitemShut {NoStop}%
\bibitem [{\citenamefont {Ashida}\ \emph {et~al.}(2020)\citenamefont {Ashida},
  \citenamefont {Imamoglu}, \citenamefont {Faist}, \citenamefont {Jaksch},
  \citenamefont {Cavalleri},\ and\ \citenamefont {Demler}}]{ashida2020}%
  \BibitemOpen
  \bibfield  {author} {\bibinfo {author} {\bibfnamefont {Y.}~\bibnamefont
  {Ashida}}, \bibinfo {author} {\bibfnamefont {A.}~\bibnamefont {Imamoglu}},
  \bibinfo {author} {\bibfnamefont {J.}~\bibnamefont {Faist}}, \bibinfo
  {author} {\bibfnamefont {D.}~\bibnamefont {Jaksch}}, \bibinfo {author}
  {\bibfnamefont {A.}~\bibnamefont {Cavalleri}}, \ and\ \bibinfo {author}
  {\bibfnamefont {E.}~\bibnamefont {Demler}},\ }\href@noop {} {\bibfield
  {journal} {\bibinfo  {journal} {arXiv preprint arXiv:2003.13695}\ } (\bibinfo
  {year} {2020})}\BibitemShut {NoStop}%
\bibitem [{\citenamefont {Schuler}\ \emph {et~al.}(2020)\citenamefont
  {Schuler}, \citenamefont {De~Bernardis}, \citenamefont {L{\"a}uchli},\ and\
  \citenamefont {Rabl}}]{schuler2020}%
  \BibitemOpen
  \bibfield  {author} {\bibinfo {author} {\bibfnamefont {M.}~\bibnamefont
  {Schuler}}, \bibinfo {author} {\bibfnamefont {D.}~\bibnamefont
  {De~Bernardis}}, \bibinfo {author} {\bibfnamefont {A.~M.}\ \bibnamefont
  {L{\"a}uchli}}, \ and\ \bibinfo {author} {\bibfnamefont {P.}~\bibnamefont
  {Rabl}},\ }\href@noop {} {\bibfield  {journal} {\bibinfo  {journal} {arXiv
  preprint arXiv:2004.13738}\ } (\bibinfo {year} {2020})}\BibitemShut {NoStop}%
\bibitem [{\citenamefont {Chakraborty}\ and\ \citenamefont
  {Piazza}(2020)}]{chakraborty2020}%
  \BibitemOpen
  \bibfield  {author} {\bibinfo {author} {\bibfnamefont {A.}~\bibnamefont
  {Chakraborty}}\ and\ \bibinfo {author} {\bibfnamefont {F.}~\bibnamefont
  {Piazza}},\ }\href@noop {} {\bibfield  {journal} {\bibinfo  {journal} {arXiv
  preprint arXiv:2008.06513}\ } (\bibinfo {year} {2020})}\BibitemShut {NoStop}%
\bibitem [{\citenamefont {Sentef}\ \emph {et~al.}(2020)\citenamefont {Sentef},
  \citenamefont {Li}, \citenamefont {K\"unzel},\ and\ \citenamefont
  {Eckstein}}]{Sentef2020}%
  \BibitemOpen
  \bibfield  {author} {\bibinfo {author} {\bibfnamefont {M.~A.}\ \bibnamefont
  {Sentef}}, \bibinfo {author} {\bibfnamefont {J.}~\bibnamefont {Li}}, \bibinfo
  {author} {\bibfnamefont {F.}~\bibnamefont {K\"unzel}}, \ and\ \bibinfo
  {author} {\bibfnamefont {M.}~\bibnamefont {Eckstein}},\ }\href {\doibase
  10.1103/PhysRevResearch.2.033033} {\bibfield  {journal} {\bibinfo  {journal}
  {Phys. Rev. Research}\ }\textbf {\bibinfo {volume} {2}},\ \bibinfo {pages}
  {033033} (\bibinfo {year} {2020})}\BibitemShut {NoStop}%
\bibitem [{\citenamefont {Kiffner}\ \emph {et~al.}(2019)\citenamefont
  {Kiffner}, \citenamefont {Coulthard}, \citenamefont {Schlawin}, \citenamefont
  {Ardavan},\ and\ \citenamefont {Jaksch}}]{Kiffner2019}%
  \BibitemOpen
  \bibfield  {author} {\bibinfo {author} {\bibfnamefont {M.}~\bibnamefont
  {Kiffner}}, \bibinfo {author} {\bibfnamefont {J.~R.}\ \bibnamefont
  {Coulthard}}, \bibinfo {author} {\bibfnamefont {F.}~\bibnamefont {Schlawin}},
  \bibinfo {author} {\bibfnamefont {A.}~\bibnamefont {Ardavan}}, \ and\
  \bibinfo {author} {\bibfnamefont {D.}~\bibnamefont {Jaksch}},\ }\href
  {\doibase 10.1103/PhysRevB.99.085116} {\bibfield  {journal} {\bibinfo
  {journal} {Phys. Rev. B}\ }\textbf {\bibinfo {volume} {99}},\ \bibinfo
  {pages} {085116} (\bibinfo {year} {2019})}\BibitemShut {NoStop}%
\bibitem [{\citenamefont {Auerbach}(2012)}]{Auerbach_book}%
  \BibitemOpen
  \bibfield  {author} {\bibinfo {author} {\bibfnamefont {A.}~\bibnamefont
  {Auerbach}},\ }\href@noop {} {\emph {\bibinfo {title} {Interacting electrons
  and quantum magnetism}}}\ (\bibinfo  {publisher} {Springer Science \&
  Business Media},\ \bibinfo {year} {2012})\BibitemShut {NoStop}%
\bibitem [{\citenamefont {Liang}\ \emph {et~al.}(1988)\citenamefont {Liang},
  \citenamefont {Doucot},\ and\ \citenamefont {Anderson}}]{Liang1988}%
  \BibitemOpen
  \bibfield  {author} {\bibinfo {author} {\bibfnamefont {S.}~\bibnamefont
  {Liang}}, \bibinfo {author} {\bibfnamefont {B.}~\bibnamefont {Doucot}}, \
  and\ \bibinfo {author} {\bibfnamefont {P.~W.}\ \bibnamefont {Anderson}},\
  }\href {\doibase 10.1103/PhysRevLett.61.365} {\bibfield  {journal} {\bibinfo
  {journal} {Phys. Rev. Lett.}\ }\textbf {\bibinfo {volume} {61}},\ \bibinfo
  {pages} {365} (\bibinfo {year} {1988})}\BibitemShut {NoStop}%
\bibitem [{\citenamefont {Arovas}\ and\ \citenamefont
  {Auerbach}(1988)}]{Arovas1988}%
  \BibitemOpen
  \bibfield  {author} {\bibinfo {author} {\bibfnamefont {D.~P.}\ \bibnamefont
  {Arovas}}\ and\ \bibinfo {author} {\bibfnamefont {A.}~\bibnamefont
  {Auerbach}},\ }\href {\doibase 10.1103/PhysRevB.38.316} {\bibfield  {journal}
  {\bibinfo  {journal} {Phys. Rev. B}\ }\textbf {\bibinfo {volume} {38}},\
  \bibinfo {pages} {316} (\bibinfo {year} {1988})}\BibitemShut {NoStop}%
\bibitem [{\citenamefont {Read}\ and\ \citenamefont
  {Sachdev}(1991)}]{Read1991}%
  \BibitemOpen
  \bibfield  {author} {\bibinfo {author} {\bibfnamefont {N.}~\bibnamefont
  {Read}}\ and\ \bibinfo {author} {\bibfnamefont {S.}~\bibnamefont {Sachdev}},\
  }\href {\doibase 10.1103/PhysRevLett.66.1773} {\bibfield  {journal} {\bibinfo
   {journal} {Phys. Rev. Lett.}\ }\textbf {\bibinfo {volume} {66}},\ \bibinfo
  {pages} {1773} (\bibinfo {year} {1991})}\BibitemShut {NoStop}%
\bibitem [{\citenamefont {Ceccatto}\ \emph {et~al.}(1993)\citenamefont
  {Ceccatto}, \citenamefont {Gazza},\ and\ \citenamefont
  {Trumper}}]{Ceccatto1993}%
  \BibitemOpen
  \bibfield  {author} {\bibinfo {author} {\bibfnamefont {H.~A.}\ \bibnamefont
  {Ceccatto}}, \bibinfo {author} {\bibfnamefont {C.~J.}\ \bibnamefont {Gazza}},
  \ and\ \bibinfo {author} {\bibfnamefont {A.~E.}\ \bibnamefont {Trumper}},\
  }\href {\doibase 10.1103/PhysRevB.47.12329} {\bibfield  {journal} {\bibinfo
  {journal} {Phys. Rev. B}\ }\textbf {\bibinfo {volume} {47}},\ \bibinfo
  {pages} {12329} (\bibinfo {year} {1993})}\BibitemShut {NoStop}%
\bibitem [{\citenamefont {Wen}(2002)}]{Wen2002}%
  \BibitemOpen
  \bibfield  {author} {\bibinfo {author} {\bibfnamefont {X.-G.}\ \bibnamefont
  {Wen}},\ }\href {\doibase 10.1103/PhysRevB.65.165113} {\bibfield  {journal}
  {\bibinfo  {journal} {Phys. Rev. B}\ }\textbf {\bibinfo {volume} {65}},\
  \bibinfo {pages} {165113} (\bibinfo {year} {2002})}\BibitemShut {NoStop}%
\bibitem [{\citenamefont {Sorella}\ \emph {et~al.}(2012)\citenamefont
  {Sorella}, \citenamefont {Otsuka},\ and\ \citenamefont {Yunoki}}]{Sorella}%
  \BibitemOpen
  \bibfield  {author} {\bibinfo {author} {\bibfnamefont {S.}~\bibnamefont
  {Sorella}}, \bibinfo {author} {\bibfnamefont {Y.}~\bibnamefont {Otsuka}}, \
  and\ \bibinfo {author} {\bibfnamefont {S.}~\bibnamefont {Yunoki}},\ }\href
  {\doibase 10.1038/srep00992} {\bibfield  {journal} {\bibinfo  {journal}
  {Scientific Reports}\ }\textbf {\bibinfo {volume} {2}},\ \bibinfo {pages}
  {992} (\bibinfo {year} {2012})}\BibitemShut {NoStop}%
\bibitem [{\citenamefont {Golinelli}\ \emph {et~al.}(1994)\citenamefont
  {Golinelli}, \citenamefont {Jolicoeur},\ and\ \citenamefont
  {Lacaze}}]{Golinelli1994}%
  \BibitemOpen
  \bibfield  {author} {\bibinfo {author} {\bibfnamefont {O.}~\bibnamefont
  {Golinelli}}, \bibinfo {author} {\bibfnamefont {T.}~\bibnamefont
  {Jolicoeur}}, \ and\ \bibinfo {author} {\bibfnamefont {R.}~\bibnamefont
  {Lacaze}},\ }\href {\doibase 10.1103/PhysRevB.50.3037} {\bibfield  {journal}
  {\bibinfo  {journal} {Phys. Rev. B}\ }\textbf {\bibinfo {volume} {50}},\
  \bibinfo {pages} {3037} (\bibinfo {year} {1994})}\BibitemShut {NoStop}%
\bibitem [{\citenamefont {Campa}\ \emph {et~al.}(2014)\citenamefont {Campa},
  \citenamefont {Dauxois}, \citenamefont {Fanelli},\ and\ \citenamefont
  {Ruffo}}]{Ruffo_book}%
  \BibitemOpen
  \bibfield  {author} {\bibinfo {author} {\bibfnamefont {A.}~\bibnamefont
  {Campa}}, \bibinfo {author} {\bibfnamefont {T.}~\bibnamefont {Dauxois}},
  \bibinfo {author} {\bibfnamefont {D.}~\bibnamefont {Fanelli}}, \ and\
  \bibinfo {author} {\bibfnamefont {S.}~\bibnamefont {Ruffo}},\ }\href@noop {}
  {\emph {\bibinfo {title} {Physics of long-range interacting systems}}}\
  (\bibinfo  {publisher} {OUP Oxford},\ \bibinfo {year} {2014})\BibitemShut
  {NoStop}%
\bibitem [{\citenamefont {Hauke}\ \emph {et~al.}(2010)\citenamefont {Hauke},
  \citenamefont {Cucchietti}, \citenamefont {Müller-Hermes}, \citenamefont
  {Ba{\~{n}}uls}, \citenamefont {Cirac},\ and\ \citenamefont
  {Lewenstein}}]{Hauke_2010}%
  \BibitemOpen
  \bibfield  {author} {\bibinfo {author} {\bibfnamefont {P.}~\bibnamefont
  {Hauke}}, \bibinfo {author} {\bibfnamefont {F.~M.}\ \bibnamefont
  {Cucchietti}}, \bibinfo {author} {\bibfnamefont {A.}~\bibnamefont
  {Müller-Hermes}}, \bibinfo {author} {\bibfnamefont {M.-C.}\ \bibnamefont
  {Ba{\~{n}}uls}}, \bibinfo {author} {\bibfnamefont {J.~I.}\ \bibnamefont
  {Cirac}}, \ and\ \bibinfo {author} {\bibfnamefont {M.}~\bibnamefont
  {Lewenstein}},\ }\href {\doibase 10.1088/1367-2630/12/11/113037} {\bibfield
  {journal} {\bibinfo  {journal} {New Journal of Physics}\ }\textbf {\bibinfo
  {volume} {12}},\ \bibinfo {pages} {113037} (\bibinfo {year}
  {2010})}\BibitemShut {NoStop}%
\bibitem [{\citenamefont {Koffel}\ \emph {et~al.}(2012)\citenamefont {Koffel},
  \citenamefont {Lewenstein},\ and\ \citenamefont {Tagliacozzo}}]{Koffel2012}%
  \BibitemOpen
  \bibfield  {author} {\bibinfo {author} {\bibfnamefont {T.}~\bibnamefont
  {Koffel}}, \bibinfo {author} {\bibfnamefont {M.}~\bibnamefont {Lewenstein}},
  \ and\ \bibinfo {author} {\bibfnamefont {L.}~\bibnamefont {Tagliacozzo}},\
  }\href {\doibase 10.1103/PhysRevLett.109.267203} {\bibfield  {journal}
  {\bibinfo  {journal} {Phys. Rev. Lett.}\ }\textbf {\bibinfo {volume} {109}},\
  \bibinfo {pages} {267203} (\bibinfo {year} {2012})}\BibitemShut {NoStop}%
\bibitem [{\citenamefont {Peter}\ \emph {et~al.}(2012)\citenamefont {Peter},
  \citenamefont {M\"uller}, \citenamefont {Wessel},\ and\ \citenamefont
  {B\"uchler}}]{Peter2012}%
  \BibitemOpen
  \bibfield  {author} {\bibinfo {author} {\bibfnamefont {D.}~\bibnamefont
  {Peter}}, \bibinfo {author} {\bibfnamefont {S.}~\bibnamefont {M\"uller}},
  \bibinfo {author} {\bibfnamefont {S.}~\bibnamefont {Wessel}}, \ and\ \bibinfo
  {author} {\bibfnamefont {H.~P.}\ \bibnamefont {B\"uchler}},\ }\href {\doibase
  10.1103/PhysRevLett.109.025303} {\bibfield  {journal} {\bibinfo  {journal}
  {Phys. Rev. Lett.}\ }\textbf {\bibinfo {volume} {109}},\ \bibinfo {pages}
  {025303} (\bibinfo {year} {2012})}\BibitemShut {NoStop}%
\bibitem [{\citenamefont {Messio}\ \emph {et~al.}(2010)\citenamefont {Messio},
  \citenamefont {C\'epas},\ and\ \citenamefont {Lhuillier}}]{Messio2010}%
  \BibitemOpen
  \bibfield  {author} {\bibinfo {author} {\bibfnamefont {L.}~\bibnamefont
  {Messio}}, \bibinfo {author} {\bibfnamefont {O.}~\bibnamefont {C\'epas}}, \
  and\ \bibinfo {author} {\bibfnamefont {C.}~\bibnamefont {Lhuillier}},\ }\href
  {\doibase 10.1103/PhysRevB.81.064428} {\bibfield  {journal} {\bibinfo
  {journal} {Phys. Rev. B}\ }\textbf {\bibinfo {volume} {81}},\ \bibinfo
  {pages} {064428} (\bibinfo {year} {2010})}\BibitemShut {NoStop}%
\bibitem [{\citenamefont {Richerme}\ \emph {et~al.}(2014)\citenamefont
  {Richerme}, \citenamefont {Gong}, \citenamefont {Lee}, \citenamefont {Senko},
  \citenamefont {Smith}, \citenamefont {Foss-Feig}, \citenamefont {Michalakis},
  \citenamefont {Gorshkov},\ and\ \citenamefont {Monroe}}]{richerme2014}%
  \BibitemOpen
  \bibfield  {author} {\bibinfo {author} {\bibfnamefont {P.}~\bibnamefont
  {Richerme}}, \bibinfo {author} {\bibfnamefont {Z.-X.}\ \bibnamefont {Gong}},
  \bibinfo {author} {\bibfnamefont {A.}~\bibnamefont {Lee}}, \bibinfo {author}
  {\bibfnamefont {C.}~\bibnamefont {Senko}}, \bibinfo {author} {\bibfnamefont
  {J.}~\bibnamefont {Smith}}, \bibinfo {author} {\bibfnamefont
  {M.}~\bibnamefont {Foss-Feig}}, \bibinfo {author} {\bibfnamefont
  {S.}~\bibnamefont {Michalakis}}, \bibinfo {author} {\bibfnamefont {A.~V.}\
  \bibnamefont {Gorshkov}}, \ and\ \bibinfo {author} {\bibfnamefont
  {C.}~\bibnamefont {Monroe}},\ }\href@noop {} {\bibfield  {journal} {\bibinfo
  {journal} {Nature}\ }\textbf {\bibinfo {volume} {511}},\ \bibinfo {pages}
  {198} (\bibinfo {year} {2014})}\BibitemShut {NoStop}%
\bibitem [{\citenamefont {Mivehvar}\ \emph {et~al.}(2021)\citenamefont
  {Mivehvar}, \citenamefont {Piazza}, \citenamefont {Donner},\ and\
  \citenamefont {Ritsch}}]{mivehvar2021cavity}%
  \BibitemOpen
  \bibfield  {author} {\bibinfo {author} {\bibfnamefont {F.}~\bibnamefont
  {Mivehvar}}, \bibinfo {author} {\bibfnamefont {F.}~\bibnamefont {Piazza}},
  \bibinfo {author} {\bibfnamefont {T.}~\bibnamefont {Donner}}, \ and\ \bibinfo
  {author} {\bibfnamefont {H.}~\bibnamefont {Ritsch}},\ }\href@noop {}
  {\bibfield  {journal} {\bibinfo  {journal} {arXiv preprint arXiv:2102.04473}\
  } (\bibinfo {year} {2021})}\BibitemShut {NoStop}%
\bibitem [{\citenamefont {Kennes}\ \emph {et~al.}(2021)\citenamefont {Kennes},
  \citenamefont {Claassen}, \citenamefont {Xian}, \citenamefont {Georges},
  \citenamefont {Millis}, \citenamefont {Hone}, \citenamefont {Dean},
  \citenamefont {Basov}, \citenamefont {Pasupathy},\ and\ \citenamefont
  {Rubio}}]{Kennes2021}%
  \BibitemOpen
  \bibfield  {author} {\bibinfo {author} {\bibfnamefont {D.~M.}\ \bibnamefont
  {Kennes}}, \bibinfo {author} {\bibfnamefont {M.}~\bibnamefont {Claassen}},
  \bibinfo {author} {\bibfnamefont {L.}~\bibnamefont {Xian}}, \bibinfo {author}
  {\bibfnamefont {A.}~\bibnamefont {Georges}}, \bibinfo {author} {\bibfnamefont
  {A.~J.}\ \bibnamefont {Millis}}, \bibinfo {author} {\bibfnamefont
  {J.}~\bibnamefont {Hone}}, \bibinfo {author} {\bibfnamefont {C.~R.}\
  \bibnamefont {Dean}}, \bibinfo {author} {\bibfnamefont {D.~N.}\ \bibnamefont
  {Basov}}, \bibinfo {author} {\bibfnamefont {A.~N.}\ \bibnamefont
  {Pasupathy}}, \ and\ \bibinfo {author} {\bibfnamefont {A.}~\bibnamefont
  {Rubio}},\ }\href {\doibase 10.1038/s41567-020-01154-3} {\bibfield  {journal}
  {\bibinfo  {journal} {Nature Physics}\ }\textbf {\bibinfo {volume} {17}},\
  \bibinfo {pages} {155} (\bibinfo {year} {2021})}\BibitemShut {NoStop}%
\bibitem [{\citenamefont {Jackeli}\ and\ \citenamefont
  {Khaliullin}(2009)}]{Jackeli2009}%
  \BibitemOpen
  \bibfield  {author} {\bibinfo {author} {\bibfnamefont {G.}~\bibnamefont
  {Jackeli}}\ and\ \bibinfo {author} {\bibfnamefont {G.}~\bibnamefont
  {Khaliullin}},\ }\href {\doibase 10.1103/PhysRevLett.102.017205} {\bibfield
  {journal} {\bibinfo  {journal} {Phys. Rev. Lett.}\ }\textbf {\bibinfo
  {volume} {102}},\ \bibinfo {pages} {017205} (\bibinfo {year}
  {2009})}\BibitemShut {NoStop}%
\bibitem [{\citenamefont {Georges}\ \emph {et~al.}(2013)\citenamefont
  {Georges}, \citenamefont {Medici},\ and\ \citenamefont
  {Mravlje}}]{Georges2013}%
  \BibitemOpen
  \bibfield  {author} {\bibinfo {author} {\bibfnamefont {A.}~\bibnamefont
  {Georges}}, \bibinfo {author} {\bibfnamefont {L.~d.}\ \bibnamefont {Medici}},
  \ and\ \bibinfo {author} {\bibfnamefont {J.}~\bibnamefont {Mravlje}},\ }\href
  {\doibase 10.1146/annurev-conmatphys-020911-125045} {\bibfield  {journal}
  {\bibinfo  {journal} {Annual Review of Condensed Matter Physics}\ }\textbf
  {\bibinfo {volume} {4}},\ \bibinfo {pages} {137} (\bibinfo {year}
  {2013})}\BibitemShut {NoStop}%
\bibitem [{\citenamefont {Rau}\ \emph {et~al.}(2016)\citenamefont {Rau},
  \citenamefont {Lee},\ and\ \citenamefont {Kee}}]{Rau2016}%
  \BibitemOpen
  \bibfield  {author} {\bibinfo {author} {\bibfnamefont {J.~G.}\ \bibnamefont
  {Rau}}, \bibinfo {author} {\bibfnamefont {E.~K.-H.}\ \bibnamefont {Lee}}, \
  and\ \bibinfo {author} {\bibfnamefont {H.-Y.}\ \bibnamefont {Kee}},\ }\href
  {\doibase 10.1146/annurev-conmatphys-031115-011319} {\bibfield  {journal}
  {\bibinfo  {journal} {Annual Review of Condensed Matter Physics}\ }\textbf
  {\bibinfo {volume} {7}},\ \bibinfo {pages} {195} (\bibinfo {year}
  {2016})}\BibitemShut {NoStop}%
\bibitem [{\citenamefont {Mivehvar}\ \emph {et~al.}(2019)\citenamefont
  {Mivehvar}, \citenamefont {Ritsch},\ and\ \citenamefont
  {Piazza}}]{Piazza2019}%
  \BibitemOpen
  \bibfield  {author} {\bibinfo {author} {\bibfnamefont {F.}~\bibnamefont
  {Mivehvar}}, \bibinfo {author} {\bibfnamefont {H.}~\bibnamefont {Ritsch}}, \
  and\ \bibinfo {author} {\bibfnamefont {F.}~\bibnamefont {Piazza}},\ }\href
  {\doibase 10.1103/PhysRevLett.122.113603} {\bibfield  {journal} {\bibinfo
  {journal} {Phys. Rev. Lett.}\ }\textbf {\bibinfo {volume} {122}},\ \bibinfo
  {pages} {113603} (\bibinfo {year} {2019})}\BibitemShut {NoStop}%
\bibitem [{\citenamefont {Joannopoulos}\ \emph {et~al.}(2008)\citenamefont
  {Joannopoulos}, \citenamefont {Johnson}, \citenamefont {Winn},\ and\
  \citenamefont {Meade}}]{joannopoulos2008}%
  \BibitemOpen
  \bibfield  {author} {\bibinfo {author} {\bibfnamefont {J.~D.}\ \bibnamefont
  {Joannopoulos}}, \bibinfo {author} {\bibfnamefont {S.~G.}\ \bibnamefont
  {Johnson}}, \bibinfo {author} {\bibfnamefont {J.~N.}\ \bibnamefont {Winn}}, \
  and\ \bibinfo {author} {\bibfnamefont {R.~D.}\ \bibnamefont {Meade}},\
  }\href@noop {} {\  (\bibinfo {year} {2008})}\BibitemShut {NoStop}%
\bibitem [{\citenamefont {Melzi}\ \emph {et~al.}(2000)\citenamefont {Melzi},
  \citenamefont {Carretta}, \citenamefont {Lascialfari}, \citenamefont
  {Mambrini}, \citenamefont {Troyer}, \citenamefont {Millet},\ and\
  \citenamefont {Mila}}]{Melzi2000}%
  \BibitemOpen
  \bibfield  {author} {\bibinfo {author} {\bibfnamefont {R.}~\bibnamefont
  {Melzi}}, \bibinfo {author} {\bibfnamefont {P.}~\bibnamefont {Carretta}},
  \bibinfo {author} {\bibfnamefont {A.}~\bibnamefont {Lascialfari}}, \bibinfo
  {author} {\bibfnamefont {M.}~\bibnamefont {Mambrini}}, \bibinfo {author}
  {\bibfnamefont {M.}~\bibnamefont {Troyer}}, \bibinfo {author} {\bibfnamefont
  {P.}~\bibnamefont {Millet}}, \ and\ \bibinfo {author} {\bibfnamefont
  {F.}~\bibnamefont {Mila}},\ }\href {\doibase 10.1103/PhysRevLett.85.1318}
  {\bibfield  {journal} {\bibinfo  {journal} {Phys. Rev. Lett.}\ }\textbf
  {\bibinfo {volume} {85}},\ \bibinfo {pages} {1318} (\bibinfo {year}
  {2000})}\BibitemShut {NoStop}%
\bibitem [{\citenamefont {Kim}\ \emph {et~al.}(2012)\citenamefont {Kim},
  \citenamefont {Casa}, \citenamefont {Upton}, \citenamefont {Gog},
  \citenamefont {Kim}, \citenamefont {Mitchell}, \citenamefont {van
  Veenendaal}, \citenamefont {Daghofer}, \citenamefont {van~den Brink},
  \citenamefont {Khaliullin},\ and\ \citenamefont {Kim}}]{Kim2012}%
  \BibitemOpen
  \bibfield  {author} {\bibinfo {author} {\bibfnamefont {J.}~\bibnamefont
  {Kim}}, \bibinfo {author} {\bibfnamefont {D.}~\bibnamefont {Casa}}, \bibinfo
  {author} {\bibfnamefont {M.~H.}\ \bibnamefont {Upton}}, \bibinfo {author}
  {\bibfnamefont {T.}~\bibnamefont {Gog}}, \bibinfo {author} {\bibfnamefont
  {Y.-J.}\ \bibnamefont {Kim}}, \bibinfo {author} {\bibfnamefont {J.~F.}\
  \bibnamefont {Mitchell}}, \bibinfo {author} {\bibfnamefont {M.}~\bibnamefont
  {van Veenendaal}}, \bibinfo {author} {\bibfnamefont {M.}~\bibnamefont
  {Daghofer}}, \bibinfo {author} {\bibfnamefont {J.}~\bibnamefont {van~den
  Brink}}, \bibinfo {author} {\bibfnamefont {G.}~\bibnamefont {Khaliullin}}, \
  and\ \bibinfo {author} {\bibfnamefont {B.~J.}\ \bibnamefont {Kim}},\ }\href
  {\doibase 10.1103/PhysRevLett.108.177003} {\bibfield  {journal} {\bibinfo
  {journal} {Phys. Rev. Lett.}\ }\textbf {\bibinfo {volume} {108}},\ \bibinfo
  {pages} {177003} (\bibinfo {year} {2012})}\BibitemShut {NoStop}%
\bibitem [{\citenamefont {Banerjee}\ \emph {et~al.}(2016)\citenamefont
  {Banerjee}, \citenamefont {Bridges}, \citenamefont {Yan}, \citenamefont
  {Aczel}, \citenamefont {Li}, \citenamefont {Stone}, \citenamefont {Granroth},
  \citenamefont {Lumsden}, \citenamefont {Yiu}, \citenamefont {Knolle} \emph
  {et~al.}}]{Banerjee2016}%
  \BibitemOpen
  \bibfield  {author} {\bibinfo {author} {\bibfnamefont {A.}~\bibnamefont
  {Banerjee}}, \bibinfo {author} {\bibfnamefont {C.}~\bibnamefont {Bridges}},
  \bibinfo {author} {\bibfnamefont {J.-Q.}\ \bibnamefont {Yan}}, \bibinfo
  {author} {\bibfnamefont {A.}~\bibnamefont {Aczel}}, \bibinfo {author}
  {\bibfnamefont {L.}~\bibnamefont {Li}}, \bibinfo {author} {\bibfnamefont
  {M.}~\bibnamefont {Stone}}, \bibinfo {author} {\bibfnamefont
  {G.}~\bibnamefont {Granroth}}, \bibinfo {author} {\bibfnamefont
  {M.}~\bibnamefont {Lumsden}}, \bibinfo {author} {\bibfnamefont
  {Y.}~\bibnamefont {Yiu}}, \bibinfo {author} {\bibfnamefont {J.}~\bibnamefont
  {Knolle}},  \emph {et~al.},\ }\href@noop {} {\bibfield  {journal} {\bibinfo
  {journal} {Nature materials}\ }\textbf {\bibinfo {volume} {15}},\ \bibinfo
  {pages} {733} (\bibinfo {year} {2016})}\BibitemShut {NoStop}%
\bibitem [{\citenamefont {Jurcevic}\ \emph {et~al.}(2014)\citenamefont
  {Jurcevic}, \citenamefont {Lanyon}, \citenamefont {Hauke}, \citenamefont
  {Hempel}, \citenamefont {Zoller}, \citenamefont {Blatt},\ and\ \citenamefont
  {Roos}}]{jurcevic2014}%
  \BibitemOpen
  \bibfield  {author} {\bibinfo {author} {\bibfnamefont {P.}~\bibnamefont
  {Jurcevic}}, \bibinfo {author} {\bibfnamefont {B.~P.}\ \bibnamefont
  {Lanyon}}, \bibinfo {author} {\bibfnamefont {P.}~\bibnamefont {Hauke}},
  \bibinfo {author} {\bibfnamefont {C.}~\bibnamefont {Hempel}}, \bibinfo
  {author} {\bibfnamefont {P.}~\bibnamefont {Zoller}}, \bibinfo {author}
  {\bibfnamefont {R.}~\bibnamefont {Blatt}}, \ and\ \bibinfo {author}
  {\bibfnamefont {C.~F.}\ \bibnamefont {Roos}},\ }\href@noop {} {\bibfield
  {journal} {\bibinfo  {journal} {Nature}\ }\textbf {\bibinfo {volume} {511}},\
  \bibinfo {pages} {202} (\bibinfo {year} {2014})}\BibitemShut {NoStop}%
\bibitem [{\citenamefont {Landini}\ \emph {et~al.}(2018)\citenamefont
  {Landini}, \citenamefont {Dogra}, \citenamefont {Kroeger}, \citenamefont
  {Hruby}, \citenamefont {Donner},\ and\ \citenamefont
  {Esslinger}}]{Landini2018}%
  \BibitemOpen
  \bibfield  {author} {\bibinfo {author} {\bibfnamefont {M.}~\bibnamefont
  {Landini}}, \bibinfo {author} {\bibfnamefont {N.}~\bibnamefont {Dogra}},
  \bibinfo {author} {\bibfnamefont {K.}~\bibnamefont {Kroeger}}, \bibinfo
  {author} {\bibfnamefont {L.}~\bibnamefont {Hruby}}, \bibinfo {author}
  {\bibfnamefont {T.}~\bibnamefont {Donner}}, \ and\ \bibinfo {author}
  {\bibfnamefont {T.}~\bibnamefont {Esslinger}},\ }\href {\doibase
  10.1103/PhysRevLett.120.223602} {\bibfield  {journal} {\bibinfo  {journal}
  {Phys. Rev. Lett.}\ }\textbf {\bibinfo {volume} {120}},\ \bibinfo {pages}
  {223602} (\bibinfo {year} {2018})}\BibitemShut {NoStop}%
\bibitem [{\citenamefont {Kroeze}\ \emph {et~al.}(2018)\citenamefont {Kroeze},
  \citenamefont {Guo}, \citenamefont {Vaidya}, \citenamefont {Keeling},\ and\
  \citenamefont {Lev}}]{Kroeze2018}%
  \BibitemOpen
  \bibfield  {author} {\bibinfo {author} {\bibfnamefont {R.~M.}\ \bibnamefont
  {Kroeze}}, \bibinfo {author} {\bibfnamefont {Y.}~\bibnamefont {Guo}},
  \bibinfo {author} {\bibfnamefont {V.~D.}\ \bibnamefont {Vaidya}}, \bibinfo
  {author} {\bibfnamefont {J.}~\bibnamefont {Keeling}}, \ and\ \bibinfo
  {author} {\bibfnamefont {B.~L.}\ \bibnamefont {Lev}},\ }\href {\doibase
  10.1103/PhysRevLett.121.163601} {\bibfield  {journal} {\bibinfo  {journal}
  {Phys. Rev. Lett.}\ }\textbf {\bibinfo {volume} {121}},\ \bibinfo {pages}
  {163601} (\bibinfo {year} {2018})}\BibitemShut {NoStop}%
\bibitem [{\citenamefont {Kre{\v{s}}i{\'c}}\ \emph {et~al.}(2018)\citenamefont
  {Kre{\v{s}}i{\'c}}, \citenamefont {Labeyrie}, \citenamefont {Robb},
  \citenamefont {Oppo}, \citenamefont {Gomes}, \citenamefont {Griffin},
  \citenamefont {Kaiser},\ and\ \citenamefont {Ackemann}}]{Krevsic2018}%
  \BibitemOpen
  \bibfield  {author} {\bibinfo {author} {\bibfnamefont {I.}~\bibnamefont
  {Kre{\v{s}}i{\'c}}}, \bibinfo {author} {\bibfnamefont {G.}~\bibnamefont
  {Labeyrie}}, \bibinfo {author} {\bibfnamefont {G.}~\bibnamefont {Robb}},
  \bibinfo {author} {\bibfnamefont {G.-L.}\ \bibnamefont {Oppo}}, \bibinfo
  {author} {\bibfnamefont {P.}~\bibnamefont {Gomes}}, \bibinfo {author}
  {\bibfnamefont {P.}~\bibnamefont {Griffin}}, \bibinfo {author} {\bibfnamefont
  {R.}~\bibnamefont {Kaiser}}, \ and\ \bibinfo {author} {\bibfnamefont
  {T.}~\bibnamefont {Ackemann}},\ }\href@noop {} {\bibfield  {journal}
  {\bibinfo  {journal} {Communications Physics}\ }\textbf {\bibinfo {volume}
  {1}},\ \bibinfo {pages} {1} (\bibinfo {year} {2018})}\BibitemShut {NoStop}%
\bibitem [{\citenamefont {Davis}\ \emph {et~al.}(2019)\citenamefont {Davis},
  \citenamefont {Bentsen}, \citenamefont {Homeier}, \citenamefont {Li},\ and\
  \citenamefont {Schleier-Smith}}]{Davis2019}%
  \BibitemOpen
  \bibfield  {author} {\bibinfo {author} {\bibfnamefont {E.~J.}\ \bibnamefont
  {Davis}}, \bibinfo {author} {\bibfnamefont {G.}~\bibnamefont {Bentsen}},
  \bibinfo {author} {\bibfnamefont {L.}~\bibnamefont {Homeier}}, \bibinfo
  {author} {\bibfnamefont {T.}~\bibnamefont {Li}}, \ and\ \bibinfo {author}
  {\bibfnamefont {M.~H.}\ \bibnamefont {Schleier-Smith}},\ }\href {\doibase
  10.1103/PhysRevLett.122.010405} {\bibfield  {journal} {\bibinfo  {journal}
  {Phys. Rev. Lett.}\ }\textbf {\bibinfo {volume} {122}},\ \bibinfo {pages}
  {010405} (\bibinfo {year} {2019})}\BibitemShut {NoStop}%
\bibitem [{\citenamefont {Gao}\ \emph {et~al.}(2020{\natexlab{b}})\citenamefont
  {Gao}, \citenamefont {Schlawin}, \citenamefont {Buzzi}, \citenamefont
  {Cavalleri},\ and\ \citenamefont {Jaksch}}]{Gao}%
  \BibitemOpen
  \bibfield  {author} {\bibinfo {author} {\bibfnamefont {H.}~\bibnamefont
  {Gao}}, \bibinfo {author} {\bibfnamefont {F.}~\bibnamefont {Schlawin}},
  \bibinfo {author} {\bibfnamefont {M.}~\bibnamefont {Buzzi}}, \bibinfo
  {author} {\bibfnamefont {A.}~\bibnamefont {Cavalleri}}, \ and\ \bibinfo
  {author} {\bibfnamefont {D.}~\bibnamefont {Jaksch}},\ }\href {\doibase
  10.1103/PhysRevLett.125.053602} {\bibfield  {journal} {\bibinfo  {journal}
  {Phys. Rev. Lett.}\ }\textbf {\bibinfo {volume} {125}},\ \bibinfo {pages}
  {053602} (\bibinfo {year} {2020}{\natexlab{b}})}\BibitemShut {NoStop}%
\bibitem [{\citenamefont {Reschke}\ \emph {et~al.}(2017)\citenamefont
  {Reschke}, \citenamefont {Mayr}, \citenamefont {Wang}, \citenamefont {Do},
  \citenamefont {Choi},\ and\ \citenamefont {Loidl}}]{Loidl}%
  \BibitemOpen
  \bibfield  {author} {\bibinfo {author} {\bibfnamefont {S.}~\bibnamefont
  {Reschke}}, \bibinfo {author} {\bibfnamefont {F.}~\bibnamefont {Mayr}},
  \bibinfo {author} {\bibfnamefont {Z.}~\bibnamefont {Wang}}, \bibinfo {author}
  {\bibfnamefont {S.-H.}\ \bibnamefont {Do}}, \bibinfo {author} {\bibfnamefont
  {K.-Y.}\ \bibnamefont {Choi}}, \ and\ \bibinfo {author} {\bibfnamefont
  {A.}~\bibnamefont {Loidl}},\ }\href {\doibase 10.1103/PhysRevB.96.165120}
  {\bibfield  {journal} {\bibinfo  {journal} {Phys. Rev. B}\ }\textbf {\bibinfo
  {volume} {96}},\ \bibinfo {pages} {165120} (\bibinfo {year}
  {2017})}\BibitemShut {NoStop}%
\bibitem [{\citenamefont {Cao}\ \emph {et~al.}(2016)\citenamefont {Cao},
  \citenamefont {Banerjee}, \citenamefont {Yan}, \citenamefont {Bridges},
  \citenamefont {Lumsden}, \citenamefont {Mandrus}, \citenamefont {Tennant},
  \citenamefont {Chakoumakos},\ and\ \citenamefont {Nagler}}]{Cao}%
  \BibitemOpen
  \bibfield  {author} {\bibinfo {author} {\bibfnamefont {H.~B.}\ \bibnamefont
  {Cao}}, \bibinfo {author} {\bibfnamefont {A.}~\bibnamefont {Banerjee}},
  \bibinfo {author} {\bibfnamefont {J.-Q.}\ \bibnamefont {Yan}}, \bibinfo
  {author} {\bibfnamefont {C.~A.}\ \bibnamefont {Bridges}}, \bibinfo {author}
  {\bibfnamefont {M.~D.}\ \bibnamefont {Lumsden}}, \bibinfo {author}
  {\bibfnamefont {D.~G.}\ \bibnamefont {Mandrus}}, \bibinfo {author}
  {\bibfnamefont {D.~A.}\ \bibnamefont {Tennant}}, \bibinfo {author}
  {\bibfnamefont {B.~C.}\ \bibnamefont {Chakoumakos}}, \ and\ \bibinfo {author}
  {\bibfnamefont {S.~E.}\ \bibnamefont {Nagler}},\ }\href {\doibase
  10.1103/PhysRevB.93.134423} {\bibfield  {journal} {\bibinfo  {journal} {Phys.
  Rev. B}\ }\textbf {\bibinfo {volume} {93}},\ \bibinfo {pages} {134423}
  (\bibinfo {year} {2016})}\BibitemShut {NoStop}%
\end{thebibliography}%

\end{document}